\documentclass[a4paper,11pt]{article}
\pdfoutput=1

\usepackage[bottom]{footmisc}

\usepackage[normalem]{ulem}
\usepackage{jheppub2}

\usepackage[dvipsnames]{xcolor}
\usepackage{lmodern}
\usepackage{amsmath}
\usepackage{amssymb}
\usepackage{enumerate}
\usepackage{mathrsfs}
\usepackage{float}
\usepackage{graphicx}
\usepackage{subcaption}
\usepackage{supertabular}
\usepackage{braket}
\usepackage{hyperref}

\definecolor{pink1}{rgb}{0.858, 0.188, 0.478}

\newcommand{\beq}{\begin{equation}}

\newcommand{\eeq}{\end{equation}}

\title{\boldmath Action, entropy and pair creation rate of charged black holes in de Sitter space}

\author[a,b]{E. Morvan,}
\author[a,b]{J.P. van der Schaar,}
\author[c]{M.R. Visser}

\affiliation[a]{Institute of Physics, University of Amsterdam, Science Park 904, PO Box 94485, 1090 GL Amsterdam, the Netherlands}
\affiliation[b]{Delta Institute for Theoretical Physics, Science Park 904, PO Box 94485, 1090 GL Amsterdam, the Netherlands}
\affiliation[c]{Department of Applied Mathematics and Theoretical Physics, University of Cambridge,
Wilberforce Road, Cambridge CB3 0WA, UK}

\emailAdd{e.k.morvanbenhaim@uva.nl}
\emailAdd{j.p.vanderschaar@uva.nl}
\emailAdd{mv551@cam.ac.uk}

\abstract{We compute and clarify the interpretation of the  on-shell Euclidean action for Reissner-Nordström black holes in de Sitter space. We show the on-shell action   is minus the sum of the   black hole   and cosmological horizon entropy for arbitrary   mass and charge in any number of dimensions. This unifying expression helps to
clear up a confusion about the Euclidean actions of extremal and ultracold black holes  in de
Sitter, as they can   be understood as  special cases of the general expression.
We then use this result to estimate the probability for the pair creation of   black holes with arbitrary mass and charge  in an empty de Sitter background, by employing the formalism of constrained instantons. Finally, we suggest that the decay of charged de Sitter black holes is governed by the gradient flow of the entropy function and that, as a consequence, the regime of light, superradiant, rapid charge emission should describe the potential decay of extreme charged Nariai black holes to singular geometries.}

\begin{document} 
\maketitle
\flushbottom

\section{Introduction}

In this work we continue our study of the Euclidean action of black holes in de Sitter space and their potential role in the partition function. In previous work \cite{SdSaction} we provided a general and covariant derivation of the Euclidean on-shell action for neutral, nonrotating black holes in de Sitter space, generalizing classic results for the Nariai instanton \cite{Ginsparg:1982rs,Bousso:1996au}, and showed it is equal to minus the sum of the two horizon entropies for arbitrary mass and in any number of dimensions    (as was  previously  done for four dimensions in  \cite{Chao:1997em, Chao:1997osu, Bousso:1998na,Gregory:2014}). We emphasized the special role of the conical singularities, as well as a proper interpretation in terms of constrained instantons (also explained in \cite{Draper:2022xzl} and first suggested in the older works \cite{Chao:1997em, Chao:1997osu}), and computed the pair production rate of an arbitrary mass black hole in de Sitter space from first principles.  

Our main motivation for studying these objects is that they might play an important role in a Euclidean path integral description of quantum gravity in de Sitter space. This is suggested by recent developments in string theory and AdS/CFT addressing black hole unitarity from a bulk perspective \cite{Penington:2019npb,Almheiri:2019psf}, which led to the discovery of (new) non-perturbative saddle solutions of the Euclidean action of Einstein gravity with negative cosmological constant and showed how to restore unitarity of black hole evaporation \cite{Penington:2019kki,Almheiri:2019qdq}. Here, we will consider non-perturbative, non-regular   saddles of the Euclidean action of Einstein gravity with positive cosmological constant and we revisit, extend and reinterpret existing results on charged black hole instantons. 

What makes de Sitter space especially interesting is the fact that the effective dynamics appears to be completely governed by the (finite) entropy of the cosmological horizon~\cite{Gibbons:1976ue, Gibbons:1977mu}, as first emphasized in \cite{Fischler-talk, Banks:2000fe, Banks:2006rx}, but also nicely spelled out in more recent work~\cite{PeningtonWitten:2022cip}. Vacuum de Sitter corresponds to the maximum entropy state, and (perturbative or non-perturbative) excited states in the effective Einstein gravity theory correspond to configurations with lower entropy, which can be described in terms of constraints on the energy (and other charges) defining the state. The probability to create such a state in the static patch of de Sitter is governed by the entropy difference and the contribution of non-perturbative states, such as black holes, to the partition function  can then be elegantly   accounted for using the constrained instanton formalism \cite{AFFLECK1981429,Cotler:2019nbi, Cotler:2020lxj}. The hope is that these results might contribute to resolving a cosmological version of the information paradox \cite{Goheer:2002vf, Aalsma:2020aib, Aalsma:2021bit, Aalsma:2022eru}, or  provide insights into the underlying microscopic holographic description of the de Sitter static patch \cite{Anninos:2011af,Nomura:2017fyh,Leuven:2018ejp,Susskind:2021dfc, Susskind:2021omt} (see also the excellent reviews~\cite{Witten:2001kn, Klemm:2004mb, Anninos:2012qw}).

As realized long ago, compared to neutral black holes in de Sitter space, the configuration space of charged black holes is quite rich, see for instance \cite{MELLOR1989361,Mellor:1989wc,Romans:1991nq, Mann:1995vb,Cardoso:2004uz}. Next to extremal black holes, bounding the maximal charge of a given mass black hole, there also exists an upper bound on the mass in a given charge sector, extending the neutral Nariai black hole to its charged analogue. Except for the special extremal, Nariai and so-called lukewarm solutions, for which the temperatures of the black hole horizon and cosmological horizon coincide,   the absence of thermal equilibrium implies that the Euclidean solution will be singular in general, a priori obscuring the meaning of the corresponding on-shell action. We  will nevertheless proceed with computing the on-shell action, since we argue the singular Euclidean solutions are meaningful as   instantons of a constrained path integral. Although (partial) results for the on-shell Euclidean action of charged de Sitter black holes   have been reported before \cite{MELLOR1989361,Mellor:1989wc,Mann:1995vb,Chao:1997osu,Cai:1997ih, Dias:2004px}, see also the recent paper~\cite{Wang:2022sbp}, in this work we will  present a completely general derivation, for any charge and mass and in any number of dimensions. Moreover, we will    correct  the on-shell action in the extremal and so-called ultracold limits. 

We will start with some preliminaries in section \ref{sec:RNdSgeometry} on charged black holes in de Sitter, setting up our conventions. We will then confirm in section \ref{sec:action} that the  action of a general Euclidean Reissner-Nordstr\"{o}m-de Sitter (RNdS) solution with conical singularities at the black hole and cosmological horizons, in arbitrary dimensions $D>3$, after imposing the (nonlinear) Smarr relation between the two horizons and including a  boundary term for the electromagnetic field, is   independent of the   Euclidean time period and is equal to 
\begin{equation} \label{eq:euclideanaction1}
    I = - \frac{\mathcal{A}_+  +\mathcal{A}_{++}}{4G}= - S_{RNdS}\,.
\end{equation}
Here $\mathcal A_+$ is the area of the (outer) black hole horizon and $\mathcal A_{++}$ is the   cosmological horizon area.  This formula suggests that the total entropy of the RNdS static patch between the two horizons is given by the sum of the horizon entropies. 
Applying this result we then compute the pair production rate of any static black hole in de Sitter space using the formalism of constrained instantons in section \ref{secCons}. We end with a comment on the Festina Lente bound \cite{Montero:2019ekk} in section \ref{wgc}, and a short summary and  open questions in   section \ref{sec:conclusion}.  
 
\section{Charged black holes in de Sitter space}
\label{sec:RNdSgeometry}

In this section we review some geometric and thermodynamic properties of nonrotating, charged black holes in a de Sitter background   (see also \cite{Tangherlini:1963bw,Mellor:1989wc,Romans:1991nq,Mann:1995vb,Astefanesei:2003gw,Cardoso:2004uz,Dolan:2013ft,Kubiznak:2015bya,Montero:2019ekk}). We mostly  keep the number of spacetime dimensions~$D$ arbitrary, but the plots are for the case $D=4.$ Specific formulas in four dimensions that are useful for making   plots can be found in Appendix \ref{4D}.

\subsection{Reissner-Nordstr\"{o}m-de Sitter geometry and phase diagram}

Charged black holes in asymptotically de Sitter spacetime  are described by the Reissner-Nordstr\"{o}m-de Sitter  solution to the Einstein-Maxwell field equation with a positive cosmological constant $\Lambda$. In static coordinates  the   RNdS line element reads
\begin{equation}
\label{eq:rndsmetric}
    ds^2 = -f(r) dt^2 + f^{-1}(r)dr^2 + r^2 d \Omega_{D-2}^2\, ,
\end{equation}
where $d \Omega_{D-2}^2$ is the metric on the round unit ($D-2$)-sphere,  and the blackening factor $f(r)$ is given by 
\begin{equation} \label{blackening1}
    f(r) = 1 - \frac{r^2}{l^2} - \frac{m}{ r^{D-3}} + \frac{q^2}{r^{2(D-3)}}\, .
\end{equation}
The length scale $l$   is the de Sitter curvature radius, which is related to the   cosmological constant by $\Lambda=(D-1)(D-2)/2l^2$.  The  mass and charge parameters, respectively $m$ and $q$, are related to the mass $M$ and electric charge $Q$    as  
\begin{equation}
\label{eq:MandQ}
    M = \frac{(D-2)\, \Omega_{D-2}}{16 \pi G}\, m\,,\qquad Q = \sqrt{\frac{(D-3)(D-2)}{8\pi G}}\, q \,,
\end{equation}
where $\Omega_{D-2}$ is the volume of the unit ($D-2$)-sphere. 
Here $M$  is   defined in a similar way as the ADM mass in asymptotically flat   space \cite{Myers:1986un}. However, unlike in asymptotically flat and  AdS space,  the asymptotic charges in de Sitter   are ``conserved in space'' rather than in time, since the Killing vector field  $\partial_t$ is spacelike near future and past infinity \cite{Ghezelbash:2001vs}. We will ignore this subtlety and refer to $M$ simply as the  ``mass''    of the black hole. Further,   we assume throughout the paper that  $m,q>0$.

\begin{figure}[t]
\centering
\begin{subfigure}{.5\textwidth}
  \centering
  \includegraphics[width=5.5cm]{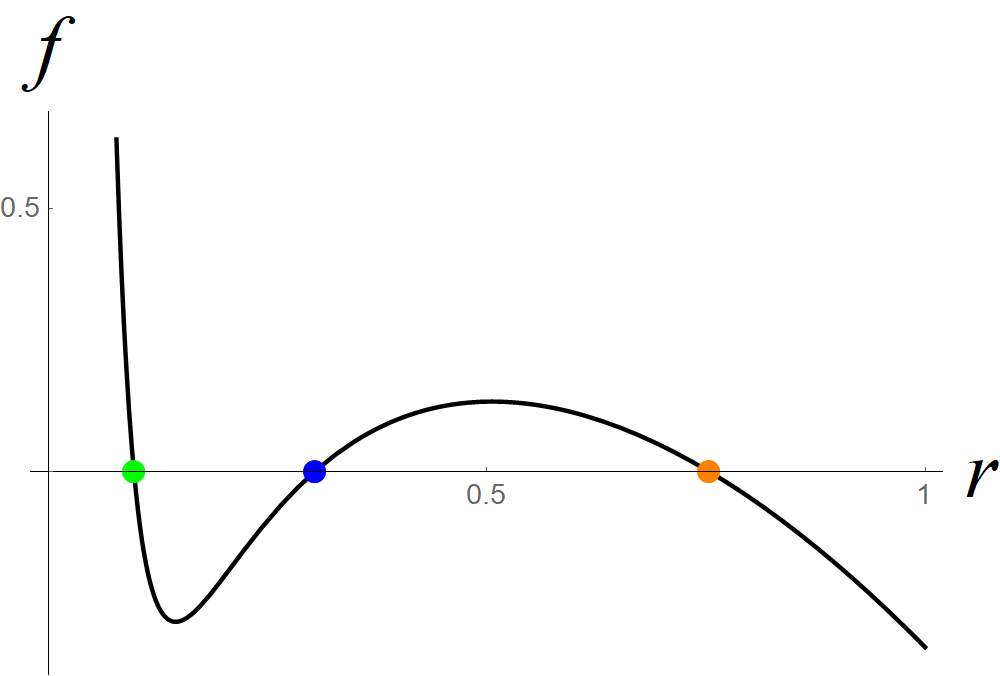}
  \caption*{a) Arbitrary $m$ and $q$}
  \label{fig:genf}
\end{subfigure}%
\begin{subfigure}{.5\textwidth}
  \centering
  \includegraphics[width=5.5cm]{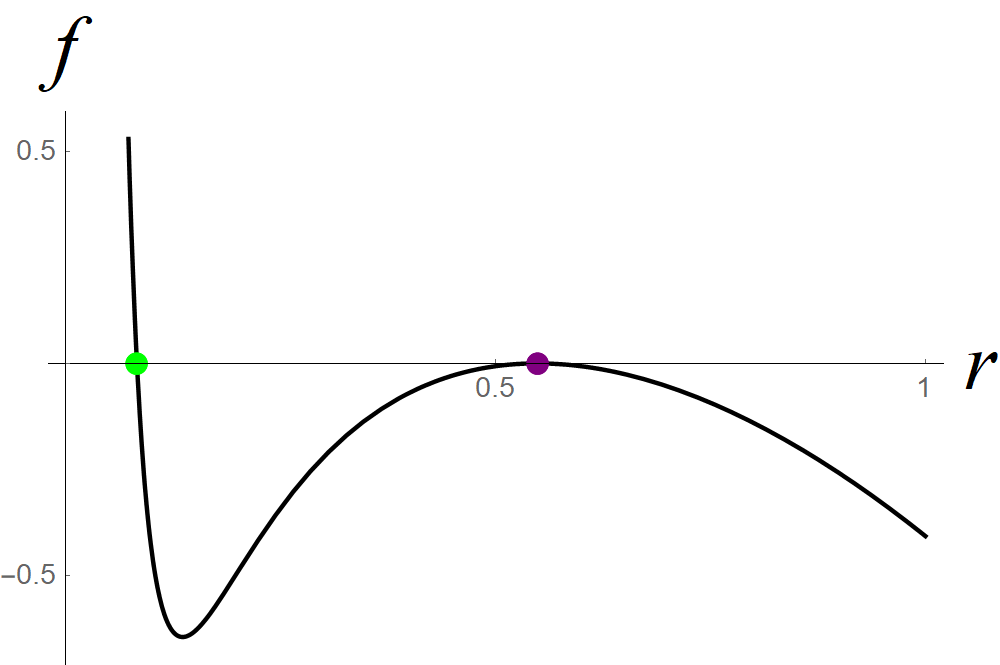}
  \caption*{b) Charged Nariai black hole}
  \label{fig:narf}
\end{subfigure}
\begin{subfigure}{.5\textwidth}
  \centering
  \includegraphics[width=5.5cm]{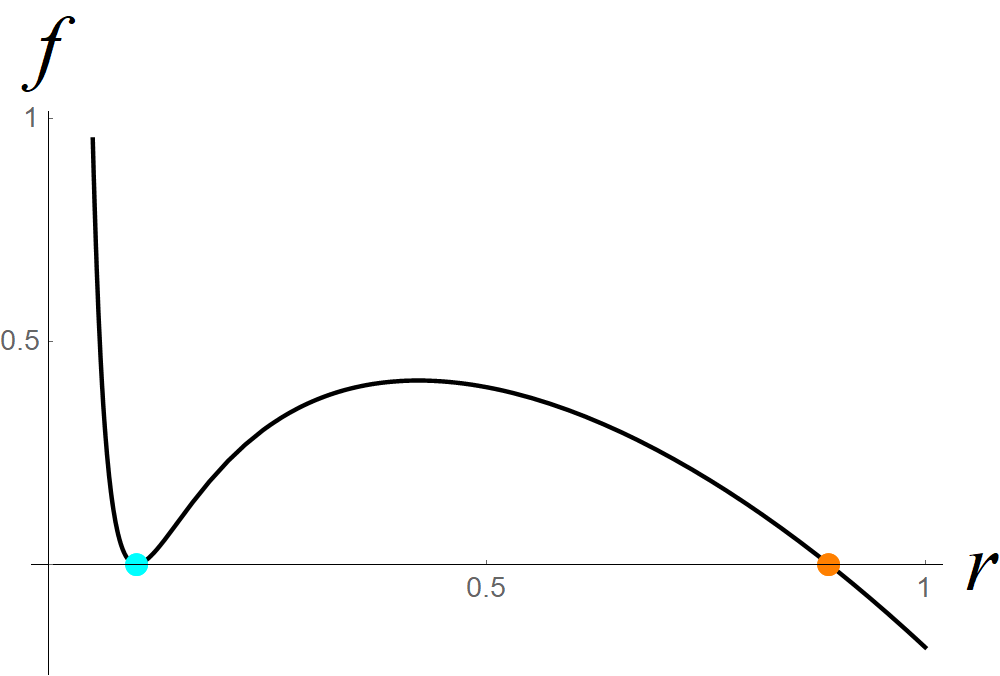}
  \caption*{c) Extremal or cold black hole}
  \label{fig:charf}
\end{subfigure}%
\begin{subfigure}{.5\textwidth}
  \centering
  \includegraphics[width=5.5cm]{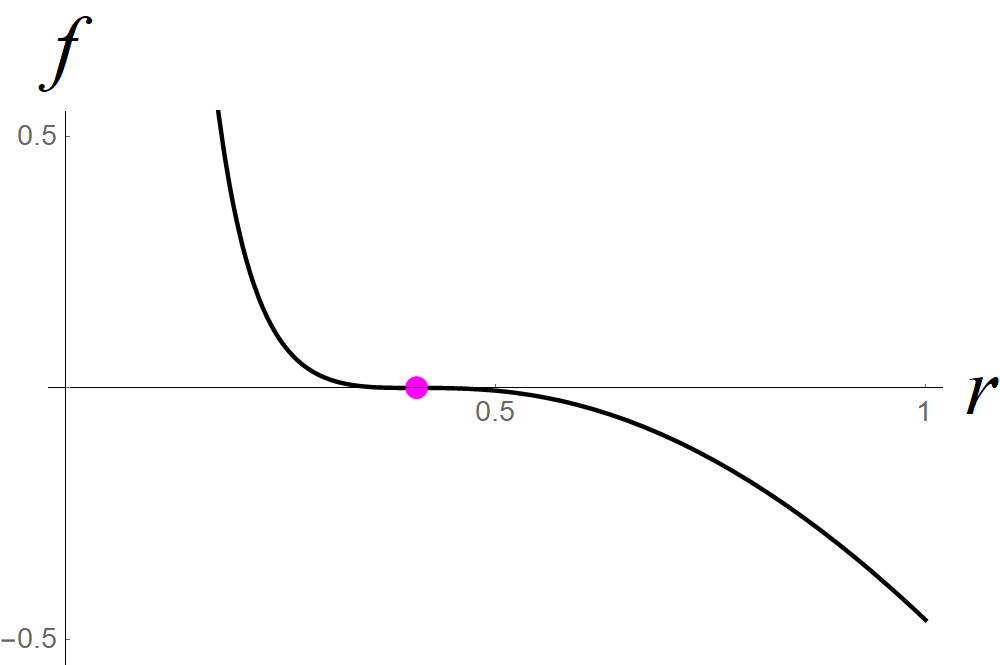}
  \caption*{d) Ultracold black hole}
  \label{fig:extf}
\end{subfigure}
\caption{Plots of the blackening factor $f$ as a function of the radius $r>0$, for a)   arbitrary $m$ and $q$, where $r_{-}\neq r_+\neq r_{++}$; b) the Nariai black hole, where   $r_{+}=r_{++}\equiv r_N$; c)   the extremal black hole,   where $r_{-}=r_{+} \equiv r_E$; d) the ultracold black hole, where   $r_{-}=r_{+}=r_{++}\equiv r_U$.}
\label{fig:Evolf}
\end{figure}

\begin{figure}[t]
\centering
\begin{subfigure}{.5\textwidth}
  \centering
  \includegraphics[width=6.5cm]{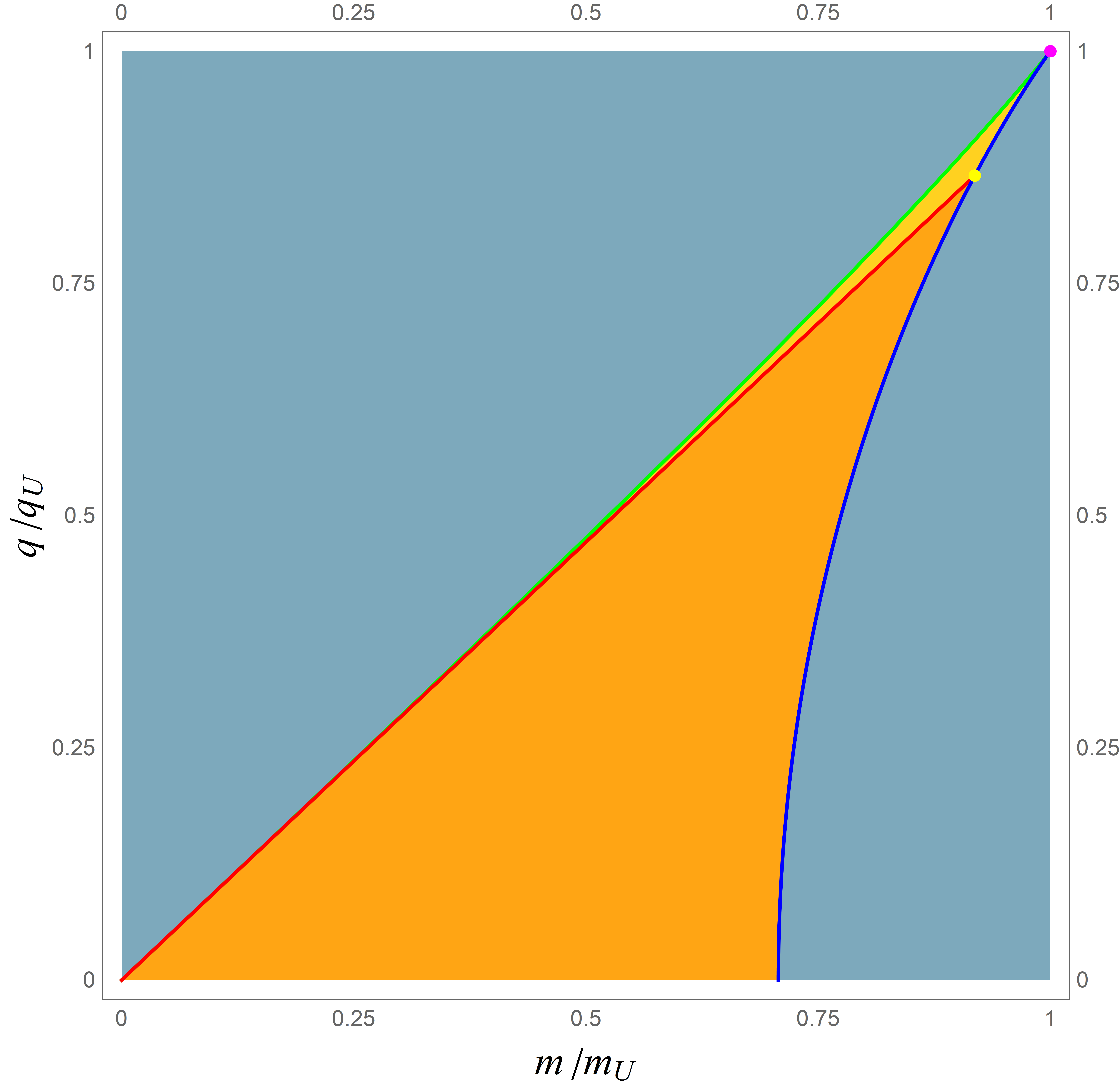}
  \caption*{a) ``Shark fin"
phase    diagram for $m/m_U$ vs. $q/q_U$}
\end{subfigure}%
\begin{subfigure}{.5\textwidth}
  \centering
  \includegraphics[width=6.5cm]{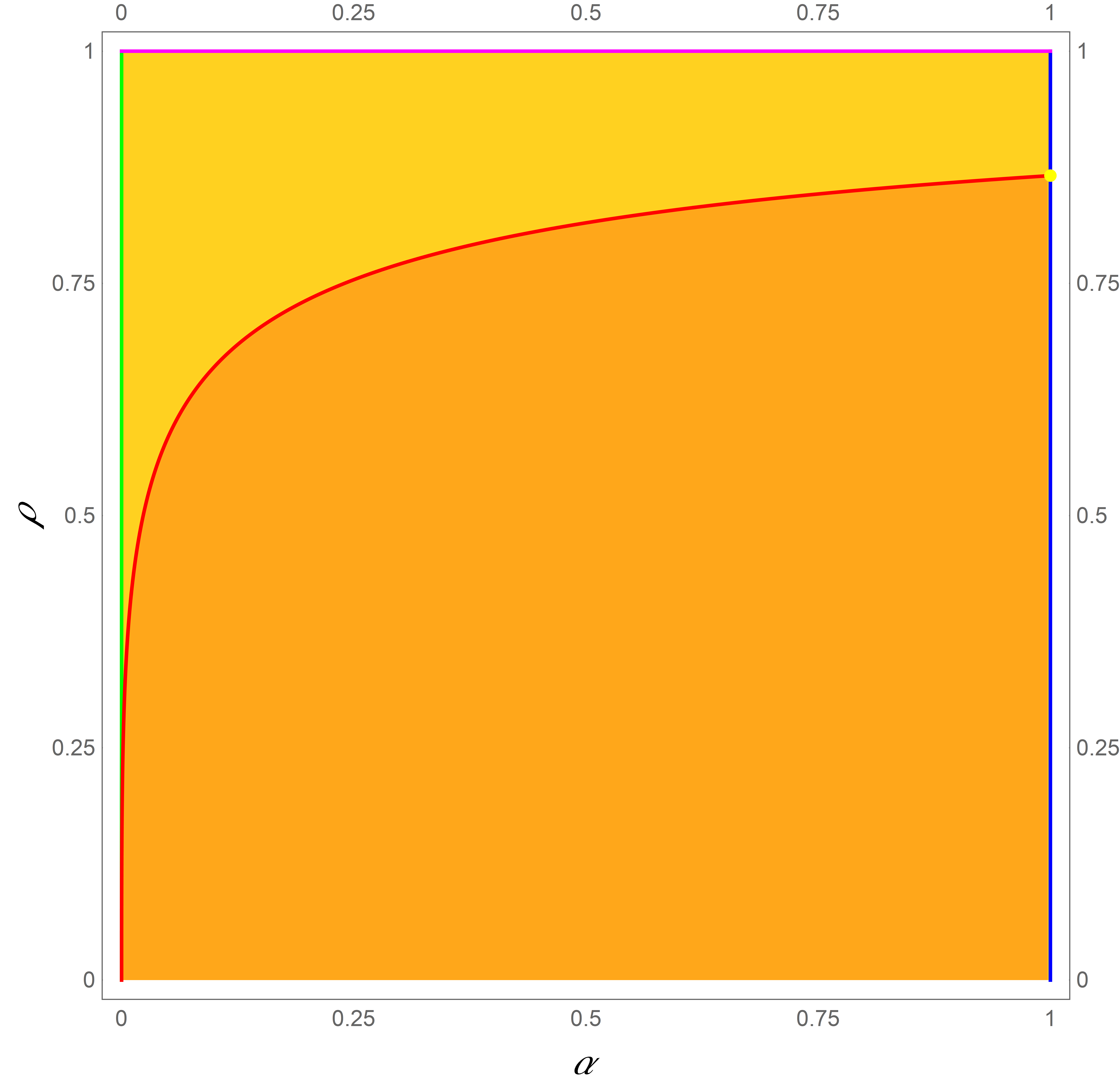}
  \caption*{b) Phase diagram for $\alpha$ vs. $\rho$.}
\end{subfigure}
\caption{Phase diagram of RNdS parameterized by $(m/m_U,q/q_U)$ (left) and  $(\alpha,\rho)$ (right). In the orange region  the bounds \eqref{bounds} are satisfied, and in the turquoise region (left)  the metric has a naked singularity. 
The  blue  boundary  corresponds to the Nariai limit,   and the  green  edge   to the extremal (cold) limit. Those boundaries meet at the magenta tip/line (left/right), associated to the ultracold limit. The red line corresponds to the lukewarm solution, for which the black hole and cosmological horizon temperatures are identical, and the yellow dot is the Nariai-lukewarm limit. In the  light orange region   black holes are colder than the cosmological horizon, whereas in the dark orange region they are hotter.}
\label{shark}
\end{figure}

The RNdS solution could be electrically or magnetically charged, but  in this paper   we  consider only the solution that is supported by an electric   gauge field  
\beq \label{Fmunu}
F = -  \frac{Q}{r^{D-2}} dt \wedge dr.
\eeq
For a fixed number of dimensions, the RNdS geometry  is fully characterized by three independent parameters $m,q$ and $l$ (or, equivalently, $M, Q$ and $\Lambda$). This implies the blackening factor $f(r)$ has four real roots, denoted by $r_{--}$, $r_-$, $r_+$ and $r_{++}$ in increasing order (for $D\ge 5$ there are additional complex roots). The negative root at $r=r_{--}$ is deemed unphysical. The   other three real roots are positive and they correspond to   horizon radii:   $r=r_-$ is the inner   black hole horizon, $r_+$ is the outer   black hole horizon, and $r_{++}$ is  the cosmological   horizon (see Figure \ref{fig:Evolf}). 
In order to avoid naked singularities    $m$ and $q$ should satisfy the following bounds  
\begin{equation} \label{bounds}
    m_E\leq m\leq m_N\,,\qquad   q_N\leq q\leq q_E  \, .
\end{equation} 
Outside this domain the roots $r_-, r_+$ and $r_{++}$ can become   complex. The left of Figure \ref{shark} shows the phase diagram of the RNdS solution in terms of $m$ and $q$, where we see that the domain~\eqref{bounds}  takes the form of a ``shark fin" \cite{Montero:2019ekk}.  Configurations inside the shark fin (orange region) have   three distinct horizons, whereas   the left and right edges of the shark fin are associated with the   extreme cases where two of the horizon radii coincide, respectively: 
 \begin{itemize}
 \item \textit{Extremal solution:}
 The lower bound on $m$ and the upper bound on $q$ correspond to the   case where the 
 inner and outer  horizon radii coincide, $r_E \equiv r_{-}=r_{+}$.  The  near-horizon topology is $AdS_2 \times S^{D-2}$ and the temperature of the extremal black hole vanishes. In the de Sitter context this  is also known as the \textit{cold} solution \cite{Romans:1991nq}. 
 \item \textit{Charged Nariai solution:} The upper bound on $m$ and the lower bound on $q$ arise from the case where the outer  horizon radius and cosmological horizon radius are the same, $r_N \equiv r_{+}=r_{++}$. The near-horizon  topology is $dS_2 \times S^{D-2}$, and the outer black hole horizon and cosmological horizon are in thermal equilibrium. 
 \end{itemize}
The mass and charge of the cold and Nariai black holes can be   found by solving $f(r_{E,N})=f'(r_{E,N})=0$, which yields the following expressions in terms of the horizon radii
\begin{equation}
\label{eq:mqbounds}
    m_{E,N}=2\left(1-\frac{D-2}{D-3}\frac{r_{E,N}^2}{l^2}\right)\,r_{E,N}^{D-3}\,,\qquad q_{E,N}^2=\left(1-\frac{D-1}{D-3}\frac{r_{E,N}^2}{l^2}\right)\,r_{E,N}^{2(D-3)} \, .
\end{equation}

\noindent  Since for any fixed mass parameter $m$, the charge parameter $q$ is bounded from above (by the cold limit), and similarly  for any fixed   $q$ there is an upper bound on   $m$ (due to Nariai limit),  there exists a de Sitter black hole with maximal mass and charge, known as: 
\begin{itemize}
\item \textit{Ultracold solution:} In this   case all three horizon radii coalesce, $r_U \equiv r_{-}=r_{+}=r_{++}$. This corresponds to the tip in the shark fin diagram where the extremal and Nariai boundaries meet. All three horizons have vanishing temperature in this case,   and the near-horizon geometry is  $M_2 \times S^{D-2},$ where $M_2$ is two-dimensional Minkowski space.  
\end{itemize}
The   horizon radius, mass and charge for the ultracold black hole are obtained by solving $f(r_U)=f'(r_U)=f''(r_U)=0$, leading to 
\begin{equation}
  r_U=\frac{D-3}{\sqrt{(D-1)(D-2)}}\,l\,,\qquad  m_U=4\frac{r_U^{D-3}}{D-1} 
    \,,\qquad q_U=\frac{r_U^{D-3}}{\sqrt{D-2}} \,.
\end{equation}
Even though the ultracold black hole has
the largest mass and charge, its (outer) horizon radius   is not maximal. The extremal horizon radius $r_E$ has the range $[0,r_U]$, whereas the Nariai horizon radius $r_N$ ranges from $r_U$ to $r_N(q=0) = l \sqrt{\frac{D-3}{D-1}}$. Hence,  the neutral Nariai black hole   has the largest outer horizon radius. 

Finally, on  the right of Figure \ref{shark} we rescaled the shark fin diagram in order to accommodate for the fact that the interval $[m_E,m_N]$ shrinks with increasing $q$. This opens up the region near the ultracold solution and makes the so-called \textit{lukewarm} solution (red line) more visible, for which
the black hole and cosmological horizon have the same temperature. We introduced a  dimensionless mass parameter $\alpha \in [0,1]$, defined via
\begin{equation}
\label{alfa}
    m=\alpha\,m_N + (1-\alpha)\,m_E \, .
\end{equation} This   allows us to  better compare and visualize thermodynamic quantities, such as the horizon temperatures and entropies. Note that  $\alpha$ is defined such that low values correspond  to small (allowed) masses, and high values to large (allowed) masses.  
Following \cite{SdSaction}, we also introduce the ratios  
\begin{equation}
     \mu\equiv m/m_N\,,\qquad \mu_E \equiv m_E/m_N\,, 
 \end{equation}
in terms of which the relation for $\alpha$ becomes
\beq
\mu= \alpha +(1-\alpha)\,\mu_E.
\eeq
Finally, for the charge we often use the dimensionless ratio
\beq
 \rho\equiv q/q_U\,,
\eeq
which ranges from $\rho=0$ (neutral black hole) to $\rho=1$ (ultracold black hole).

\subsection{Thermodynamics: horizon temperatures, first law and  Smarr formula \label{temps}}
Next we briefly discuss the thermodynamics of the Reissner-Nordstr\"{o}m-de Sitter spacetime. Since RNdS has three event horizons, there are three associated temperatures and entropies  
\beq
\label{temp}
T_{h}=\frac{\kappa_{h}}{2\pi}\,, \qquad S_{h}=\frac{\mathcal{A}_{h}}{4G}\,, \qquad \text{with}\quad h=\{-,+,++\}\, ,
\eeq
where $\kappa$ denotes the surface gravity and $\mathcal{A}$  the horizon area. For general values of the mass and charge,  the surface gravities of the three horizons are different, hence the horizons are not in thermodynamic equilibrium. However, as we will discuss in more detail below, for specific values of $m$ and $q$ (two or three of)  the surface gravities coincide, hence the associated horizon temperatures are the same.
We will focus on the thermodynamics of the static patch between the outer black hole horizon and the cosmological horizon $r\in [r_+,r_{++}]$, since that is the region captured by the Euclidean RNdS geometry.

\subsubsection*{Horizon temperatures}

\begin{figure}[t]
\centering
\begin{subfigure}{.5\textwidth}
  \centering
  \includegraphics[width=6.5cm]{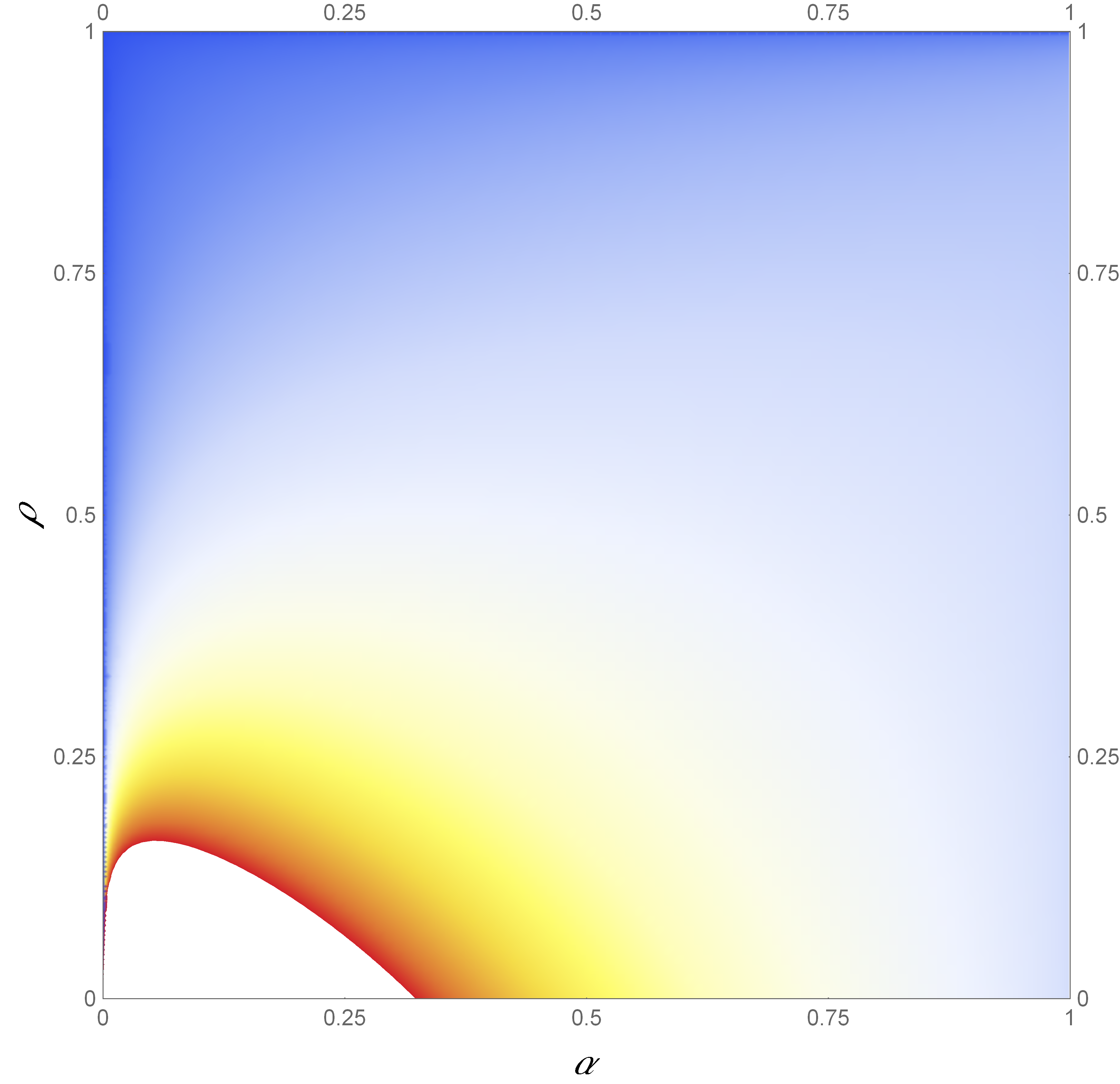}
  \caption*{a) Black hole temperature $T_{+}$}
\end{subfigure}%
\begin{subfigure}{.5\textwidth}
  \centering
  \includegraphics[width=6.5cm]{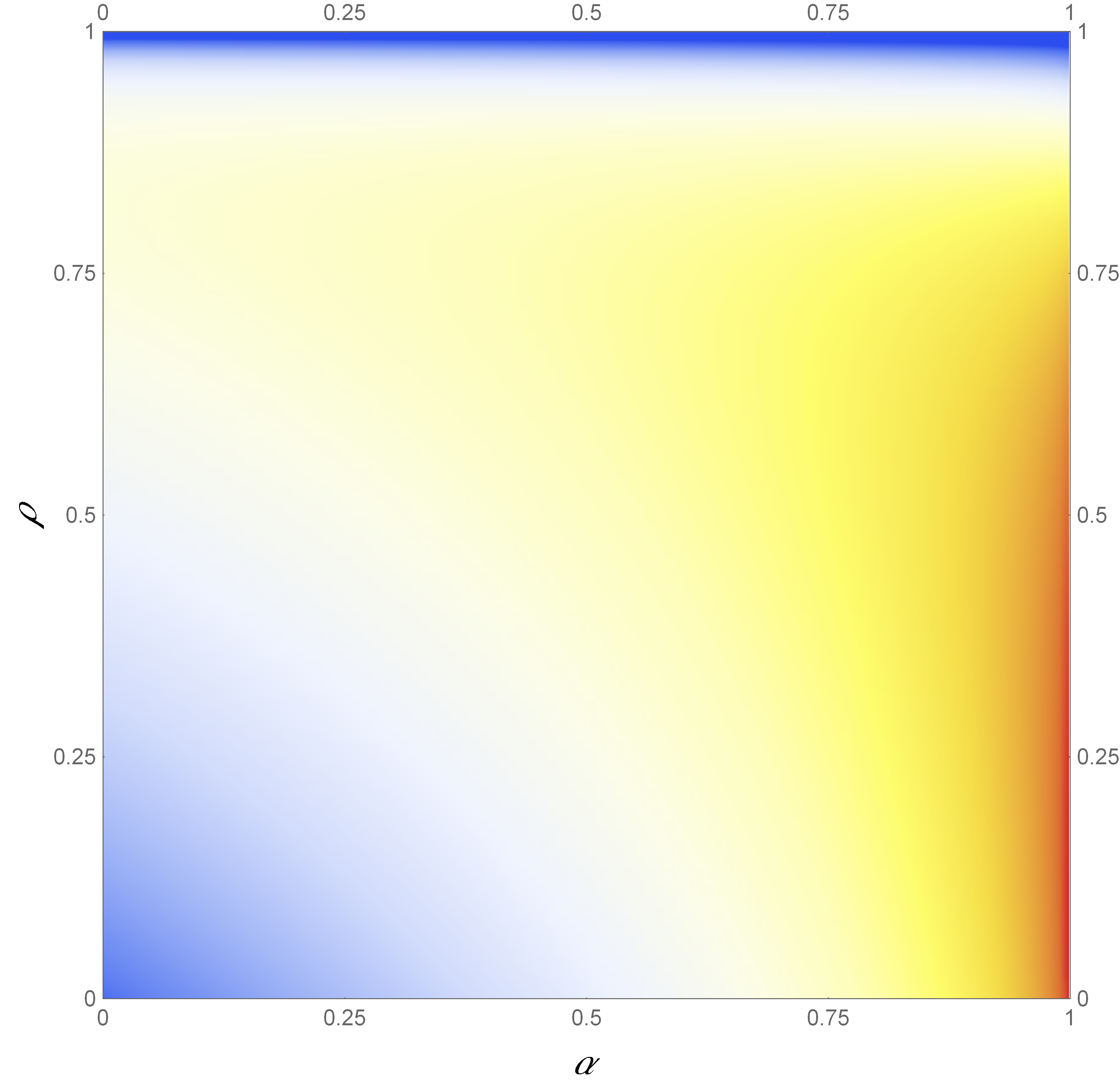}
  \caption*{b)  Cosmological horizon temperature $T_{++}$}
\end{subfigure}
\caption{Heat maps in terms of $(\alpha,\rho)$ for the black hole temperature (left) and the cosmological horizon temperature (right). For low mass and charge (in the white region at the bottom left corner)   the black hole is extremely hot. }
\label{fig:phasediatemps}
\end{figure}

The horizon temperature depends on the normalization of the horizon generating Killing vector $\xi$, which cannot be fixed at the asymptotic boundary in de Sitter space, as is usually done for Schwarzschild spacetime.
 Bousso and Hawking \cite{Bousso:1996au} normalized the timelike Killing vector of Schwarzschild-de Sitter spacetime instead on  the geodesic at constant angular variables  between the black hole and the cosmological horizon. The geodesic  lies at a particular radius $r_{\mathcal O}$ where an observer can stay in place without accelerating, i.e., the gravitational pull from the black hole and cosmological acceleration balance out. The blackening factor attains a maximum at this radius,  $f'(r_{\mathcal O})=0$, and the Killing vector is normalized such that $\xi^2 = -1$ at $r=r_{\mathcal O}$. In the limit to ($\Lambda=0$) Schwarzschild spacetime the radius goes to infinity, and in the limit to empty  ($M=0$) de Sitter spacetime the radius goes to zero. Hence, the radius $r_{\mathcal O}$ agrees in these limits with the standard location where the Killing vector is normalized.

In RNdS, the equation $f'(r_{\mathcal O})=0$ has two positive roots: a local minimum between the inner and outer horizon, and a local maximum between the outer horizon and cosmological horizon.  We normalize the Killing vector on the local maximum to mimic the normalization for the timelike Killing vector of  Schwarzschild-de Sitter.  
The blackening factor evaluated at $r_{\mathcal O}$ is given by
\begin{equation}
    f(r_{\mathcal{O}})=1-\frac{D-1}{D-3}\frac{r_{\mathcal{O}}^2}{l^2}-\frac{q^2}{r_{\mathcal{O}}^{2(D-3)}}\, .
\end{equation}
The   normalized horizon generating Killing vector is $\xi =    \partial_t /\sqrt{f(r_{\mathcal O})}$ and the surface gravity defined with respect to this Killing vector is given by  $\kappa_h=f'(r_h)/(2\sqrt{f(r_{\mathcal{O}})})$. Hence, for the RNdS solution the   temperature of the different horizons takes the following form  
\begin{align}
\label{eq:temps}
   T_{h}= \frac{D-3}{4\pi\,r_{h}\sqrt{f(r_{\mathcal{O}})}}\left|1-\frac{D-1}{D-3 }\frac{r_{h}^2}{l^2}-\frac{q^2}{r_{h}^{2(D-3)}}\right|
    \,.
\end{align}
We will now discuss several special cases for which the    horizon temperatures are the same. 
The normalization with respect to the geodesic observer between $r_+$ and $r_{++}$ is particularly appropriate for the Nariai black hole, since  the outer horizon and cosmological horizon temperatures remain finite in this limit, which is expected since the near-horizon Nariai geometry is two-dimensional pure de Sitter space times a sphere.   (The temperatures $T_+$ and $T_{++}$ defined with respect to the Killing vector $\partial_t$  seem inappropriate for the Nariai solution since they vanishes in that limit.)
In the Nariai case the black hole and cosmological horizons have the same radius as the geodesic, i.e., $r_+=r_{++}=r_{\mathcal{O}}=r_N$. Moreover, the temperatures of the black hole and cosmological horizon  are the same, and they  equal to 
\begin{equation}
\label{t2eqigen}
    T_N=\frac{\sqrt{D-3}}{2\pi\,r_N}\sqrt{1-(D-2)\,\frac{q^2}{r_N^{2(D-3)}}}\, . 
\end{equation}
For   the critical values $q=0$ and $q=q_U$ the Nariai temperature becomes, respectively,
\begin{equation} \label{ultracoldtemp}
T_{N}(q=0)=\frac{\sqrt{D-1}}{2\pi\,l}\,,
    \qquad T_U \equiv T_N(q=q_U) =0\,.
\end{equation}
The former is the temperature of the neutral Nariai black hole, which agrees with previous findings \cite{Bousso:1996au,Svesko:2022txo,SdSaction}, and the latter is the temperature of the ultracold solution   which vanishes for all three horizons.  

Furthermore, in the extremal (cold) case the  temperatures of the  outer and inner  horizons  vanish,\footnote{This is not the case if the temperatures is normalized with respect to the other positive root $\Tilde{r}_{\mathcal{O}}$ of $f'(r)=0$, where the blackening factor has a local minimum, located between $r_-$ and $r_+$. For this other normalization the black hole temperature does not go to zero in the extremal limit, $\Tilde{T}_{E,-}=\Tilde{T}_{E,+}\neq0$, whereas the Nariai black hole  temperature does vanish, $\Tilde{T}_{N,+}=\Tilde{T}_{N,++}=0$, and the temperature of the ultracold black hole as well, $\tilde{T}_U=0$.}   
$
    T_{E,+}=0\, ,
$
and the temperature of the cosmological horizon is finite   and  given by
\begin{equation}
\label{t2eqigen2}
    T_{E,++}=\frac{3-D}{4\pi\,r_{E,++}\sqrt{f(r_{\mathcal{O}})}}\left(1-\frac{D-1}{D-3} \frac{r_{E,++}^2}{l^2}  -\frac{q^2}{r_{E,++}^{2(D-3)}}\right)\, , 
\end{equation}
where $r_{E,++}$ stands for the cosmological horizon  radius of the extremal solution. For zero charge $q=0$ ($r_{E,++}=l$) the temperature reduces to the empty de Sitter temperature $T_{dS}$ (\ref{eq:Tds}) and for $q=q_U$ ($r_{E,++}=r_U$) the temperature vanishes, cf.   \eqref{ultracoldtemp}.

\begin{figure}[t]
\centering
\begin{subfigure}{.5\textwidth}
  \centering
  \includegraphics[width=6.5cm]{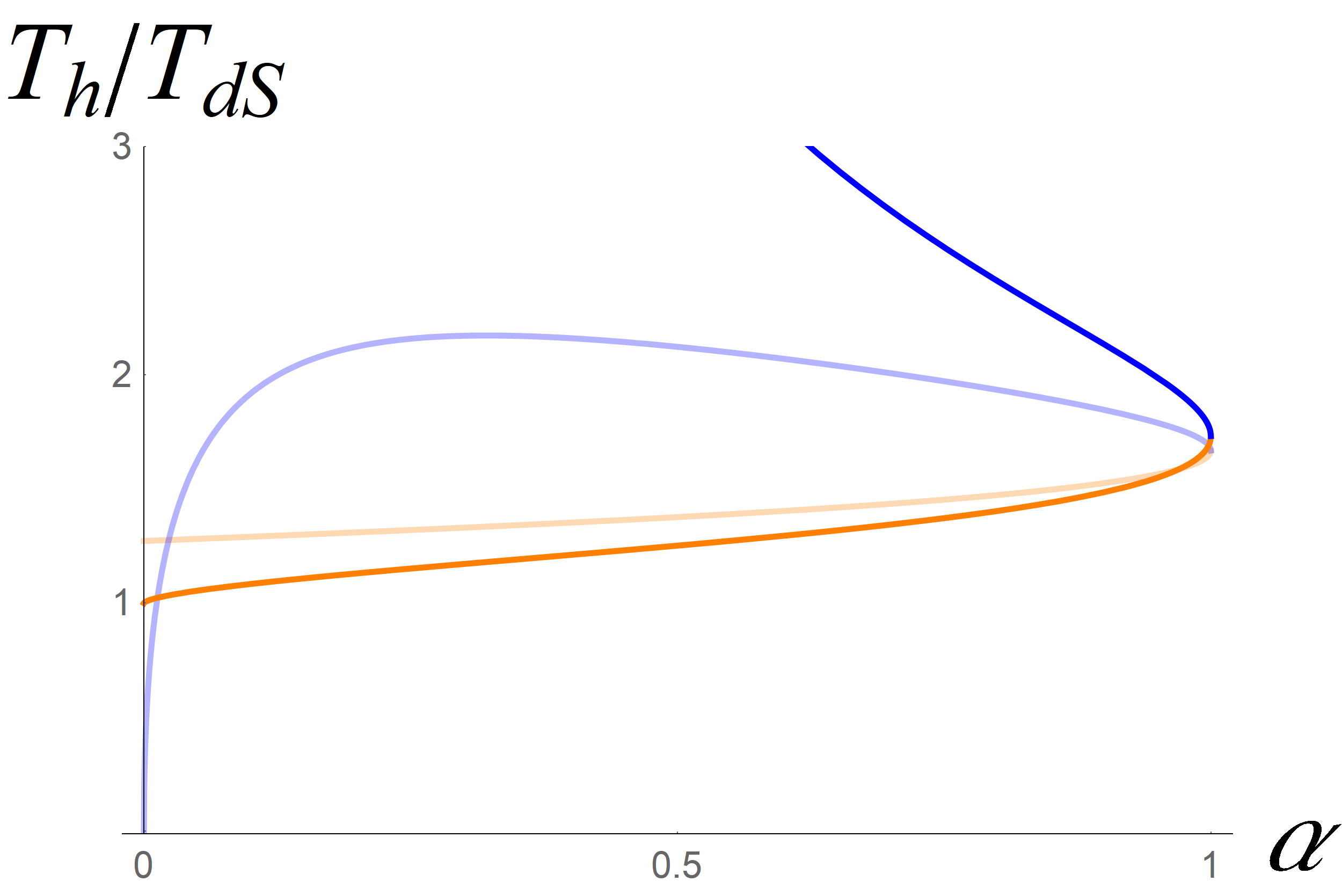}
  \caption*{a) $\rho=0,1/2$.}
\end{subfigure}%
\begin{subfigure}{.5\textwidth}
  \centering
  \includegraphics[width=6.5cm]{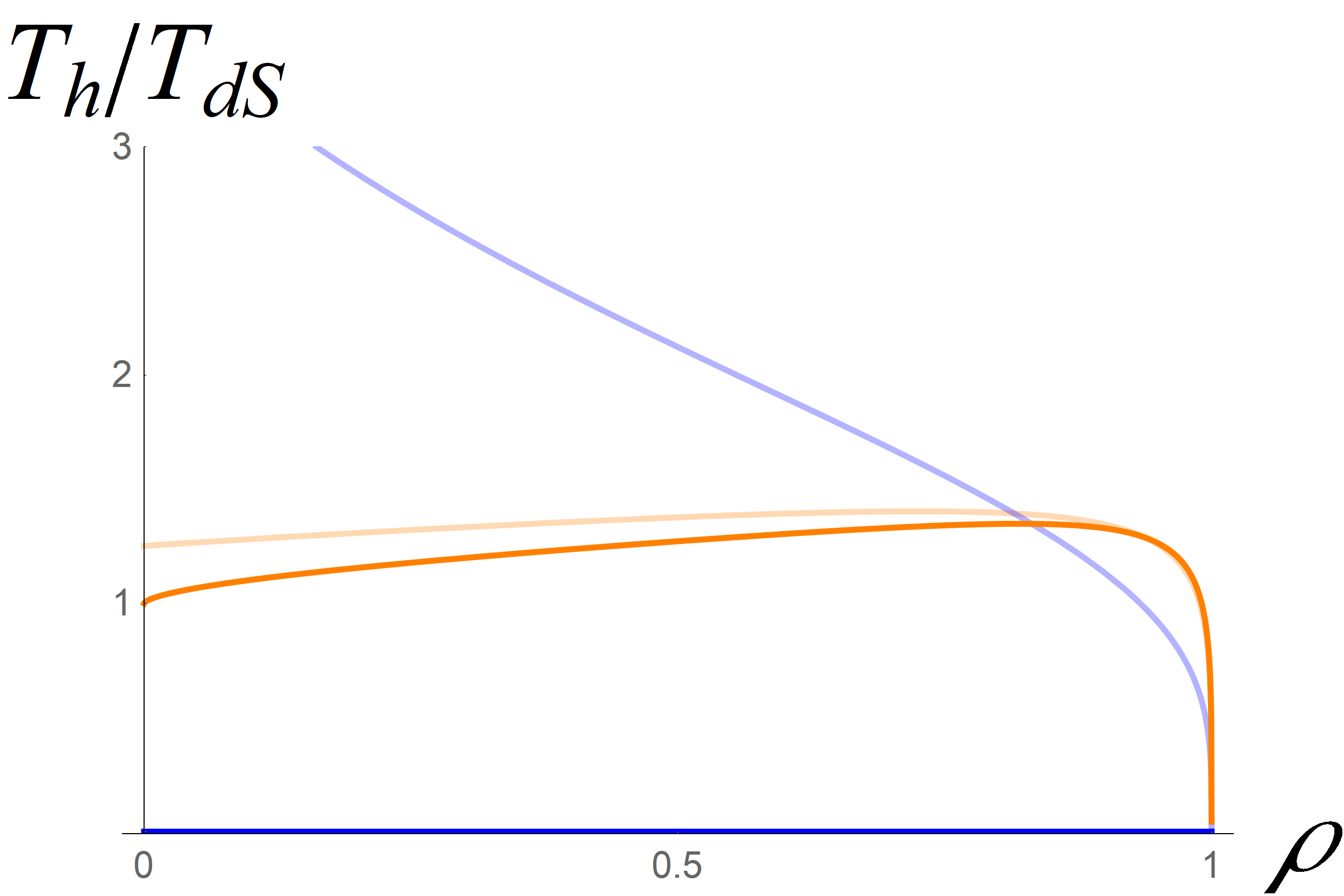}
  \caption*{b) $\alpha=0,1/2$}
\end{subfigure}
\caption{Horizon temperatures (a) as a function of the mass parameter $\alpha$ for fixed $\rho=0,1/2$ and (b) as a function of the charge parameter $\rho$ for $\alpha=0,1/2$. The black hole temperature $T_{+}$ is shown in blue and the cosmological horizon temperature $T_{++}$ in orange.  Dark  colors correspond to $\alpha,\rho=0$, and light  colors to $\alpha,\rho=1/2.$ }
\label{fig:RNNtempshades}
\end{figure}

Besides the Nariai solution there is  another thermal equilibrium solution for which the (outer) black hole horizon and cosmological horizon temperature are the same. This is the so-called \emph{lukewarm} solution \cite{Romans:1991nq}.
In Figure \ref{fig:RNNtempshades} we show that   the black hole temperature $T_{+}$ and the cosmological horizon temperature $T_{++}$ cross for specific combinations of $\alpha$ and $\rho$, or equivalently $m$ and $q$, which corresponds   to the lukewarm case. 
 When solving for $T_{++}=T_{+}$, one  finds that the  mass and charge parameter are related by $m_L=2 q_L$ in $D=4$. In higher dimensions the relation between the (lukewarm) mass and charge is more complicated, since   it depends non-trivially on the horizon radii, cf. equation \eqref{eq:mqel}.  Further, the    temperature of the lukewarm black hole, $  T_+=T_{++}(\equiv T_L)$, can be expressed in general dimensions as a function of the outer horizon radius $  r_{L,+}$ and the cosmological horizon radius $ r_{L,++}$  
\begin{equation}
   T_{L}=\frac{1}{2\pi\sqrt{f(r_{\mathcal{O}})}\, l^2}\left[r_{L,++}+r_{L,+}^{D-2}(r_{L,+}+r_{L,++})\left(\frac{D-3}{r_{L,+}^{D-2}-r_{L,+}r_{L,++}^{D-3}}+\frac{D-2}{r_{L,+}^{D-2}-r_{L,++}^{D-2}}\right)\right]\,.
\end{equation}
For $q=0$ and $q=q_{N}$ this reduces to, respectively,
\begin{equation}
\label{eq:Tds}
  T_{dS} \equiv  T_{L}(q=0)=\frac{1}{2\pi l} \,,\qquad T_{LN} \equiv T_{L}(q=q_{N}) =\frac{\sqrt{D-1}}{\sqrt{6}\pi l}\, .
\end{equation}
The former being the empty de Sitter temperature and the latter the lukewarm Nariai temperature of the black hole and cosmological horizon.


\subsubsection*{First law and generalized Smarr formula}  
The horizon temperatures, entropies,   and charge are related by the   first law for the static patch between  the black hole and cosmological horizon
(see, e.g., \cite{Dolan:2013ft})
\beq
\label{eq:firstlaw}
T_+ \delta S_{+} + T_{++} \delta S_{++} + (\Phi_{+} - \Phi_{++}) \delta Q=0\, . 
\eeq
Here $\Phi_+$ and $\Phi_{++}$  are the static electric potentials evaluated at the black hole and cosmological horizon, defined with respect to the same Killing vector normalization as the temperatures,\footnote{This normalization will cause the electric potentials to diverge in the Nariai limit, but since $\Phi_+=\Phi_{++}$ in that limit that will not affect any of our results. We thank Alejandra Castro for making us aware of this.}
\begin{equation} \label{potentials}
    \Phi_{+,++} = - A_\mu \xi^\mu \Big |_{r=r_{+,++}} = \frac{Q}{(D-3) r_{+,++}^{D-3} \, \sqrt{f(r_\mathcal{O}})} \,.
\end{equation}
Without variations the thermodynamic quantities satisfy the (generalized) Smarr formula
\begin{equation}
\label{eq:rndssmarr}
    T_+ S_+  +  T_{++} S_{++}  + \frac{D-3}{D-2} (\Phi_{+} - \Phi_{++})Q   - \frac{\Theta \Lambda}{(D-2)4\pi G} =0 \,.
\end{equation}
The quantity $\Theta$ is the ``thermodynamic volume'' or   ``Killing volume'',   defined as
\cite{Jacobson:2018ahi} \begin{equation} \label{eq:killingvolume}
    \Theta \equiv \int_\Sigma dV \sqrt{-\xi^2}\,, 
\end{equation}
where $\Sigma$ is the spatial slice between the outer horizon and cosmic horizon,  
and $\xi $ is the time translation generating Killing vector, normalized as $\xi = 
\partial_t / \sqrt{f(r_{\mathcal O})}$.

\section{Euclidean action   of Reissner-Nordstr\"{o}m-de Sitter space}
\label{sec:action}

\subsection{Hartle-Hawking wavefunction and instantons}

We are interested in computing the pair creation rate of charged black holes in a de Sitter background for any 
mass and charge.\footnote{From a global de Sitter perspective this is naturally interpreted as creating a pair of opposite charge, opposite mass, black holes, due to the opposite time orientation in two conjugate static patches. In a single static patch region this instead corresponds to the spontaneous creation of a single charged black hole.} The pair creation of such  black holes is described by the propagation from nothing to a spatial surface $\Sigma$ with topology $S^{D-2} \times S^{1}$ for generic mass and charge (and $\Sigma$ has topology $S^{D-2}\times R$ for   extreme and ultracold black holes). Given certain boundary conditions on $\Sigma$ fixing the mass and charge, the probability amplitude for this process is given by the no-boundary Hartle-Hawking wavefunction, which can be formally represented  as a Euclidean path integral   \cite{Hawking:1995ap,Mann:1995vb}
\begin{equation}
    \Psi_\Sigma   = \int_{\varnothing}^{\Sigma} [d g] [dA] e^{-I_{1/2}[g,A]}\,. 
\end{equation}
The integral is over all metrics and gauge fields that agree with the boundary data on~$\Sigma$,  and $I_{1/2}$ is the   action of a Euclidean manifold with boundary $\Sigma$. If there exists a Euclidean saddle solution satisfying the boundary conditions, then in the saddle point approximation the wavefunction is estimated by $\Psi_\Sigma \approx e^{-I_{1/2}}$, where $I_{1/2}$ is the on-shell action of the instanton and $\Sigma$ should be viewed as labeling the different final states. Moreover, the pair creation rate of fixed charge and mass black holes is approximated by the square of the amplitude, i.e.,  \begin{equation} \label{squareofHH}
    \Gamma_\Sigma = \Psi^*_\Sigma \Psi_\Sigma  \approx e^{-2 I_{1/2}}.
\end{equation}
We will be interested, however, in the probability to produce a pair of \emph{arbitrary} mass and charge black holes. We will argue in the next section that this is computed by the full gravitational path integral (without any boundary conditions) which can be written, in a saddle point approximation, as an integral over the mass and charge
\begin{equation} \label{pathintegralHH}
\int d\Sigma \, \Psi^*_\Sigma \Psi_\Sigma = \int [d g] [dA] e^{-I[g,A]} \approx \int d\alpha  d \rho \exp[- I]
\end{equation}
 Here $d\Sigma$ should be understood as the (formal) measure over the complete set of final states at the original boundary, and $I[g,A]$ is the Euclidean action on the entire (compact) manifold. In the saddle point approximation we introduced $I=2 I_{1/2}$, corresponding to the action of a ``bounce'' Euclidean solution (the entire closed manifold).  In this section we compute the on-shell action of  these ``bounce'' Euclidean solutions, which is twice the action of a standard instanton with boundary $\Sigma$, and in the next section we will calculate the pair creation rate.  
 
The Euclidean RNdS solutions provide the instantons for the pair creation of charged de Sitter black holes. 
Their metric can be obtained from the Lorentzian metric by analytically continuing $t \to it= \tau$,  
\begin{equation}
    ds^2 = f(r) d\tau^2 + f^{-1}(r) dr^2 + r^2 d\Omega_{D-2}^2,
\end{equation}
where $f(r)$ is given by \eqref{blackening1} and the Euclidean time $\tau$ is periodically identified with  inverse temperature $\beta$. For arbitrary $\beta$ the Euclidean RNdS spacetime is regular everywhere except at the black hole and cosmological horizon, where there are conical singularities. One of the conical singularities can be removed by choosing $\beta$ appropriately. However, it is not possible to remove both conical singularities at the same time, except in the extreme cases discussed in the previous section. Indeed, the extremal, Nariai, cold and lukewarm Euclidean spacetimes are smooth everywhere, whereas for a generic mass and charge the Euclidean RNdS geometry is singular at   (at least one of) the Euclidean horizons. The extreme RNdS solutions are, therefore, proper instantons of the Euclidean path integral, but we will argue in     section \ref{secCons} that generic Euclidean RNdS geometries are   instantons of a constrained path integral where the mass and charge are kept fixed.  

\subsection{On-shell Euclidean action}

In order to compute the on-shell action of the entire Euclidean RNdS section  $\mathcal M$ 
   we cut the compact geometry in half. We first calculate the action of the two ``caps'' separately, denoted by $\mathcal M_{1/2}$, and then sum the results to obtain the action of the entire Euclidean section.    This cutting and pasting procedure is consistent with the path integral representation in \eqref{pathintegralHH}. Moreover, it seems necessary due to a subtlety related to the boundary term of the Maxwell action, which localizes on the conical singularities, and does not vanish when the two caps are glued together. We cut the Euclidean section along   the spatial slice that has zero extrinsic curvature, which corresponds to the standard choice of boundary $\Sigma$ for the Hartle-Hawking wavefunction. 

The appropriate   off-shell action of half the Euclidean section $\mathcal M_{1/2}$,  for boundary data that fix the induced metric  and electric charge on $\Sigma$,  is given by \cite{Braden:1990hw,Hawking:1995ap}
\begin{equation} \label{startingaction}
   I_{1/2}=  -\int_{\mathcal{M}_{1/2}} \!\!d^D x \sqrt{g} \left[\frac{1}{16\pi G}(R-2\Lambda)-\frac{1}{4}F^2\right]- \int_{\Sigma} d^{D-1} x \sqrt{h} F^{\mu\nu} n_\mu A_\nu\, .
\end{equation}
Here $n_\mu$ is the unit normal to the boundary $\Sigma$ and $h$ is the determinant of the induced metric on $\Sigma$. We did not include the Gibbons-Hawking-York boundary term, since it vanishes on a   surface $\Sigma$ that has zero extrinsic curvature.  We want to evaluate the action on half of the Euclidean   RNdS geometry where  the Euclidean time has the range   $\tau \in [0,\beta/2]$, for an arbitrary inverse temperature $\beta$.    The Ricci scalar of the RNdS geometry has a regular part and a divergent part   
\beq
  R=  R_{{bulk}}+  R_{{con}}\,,
\eeq
where the second part is due to the conical singularities at the two Euclidean horizons and is given by
\beq
R_{con} =  4\pi \delta (r=r_{+}) + 4 \pi \delta (r=r_{++})\, .
\eeq 
To begin with, we compute   the   contribution  from the regular part of the Ricci scalar to the on-shell Euclidean action. Away from the conical singularities the RNdS spacetime satisfies the Einstein equation 
\begin{equation}
    R_{\mu \nu} - 2 \Lambda g_{\mu \nu}= 8 \pi G \left ( T_{\mu \nu} - \frac{T}{D-2}g_{\mu \nu}\right) \,.
\end{equation}
For the action \eqref{startingaction} the  Maxwell energy-momentum tensor  is given by
\beq
T_{\mu \nu}  = F_{\mu \alpha} {F_\nu}^\alpha - \frac{1}{4}g_{\mu \nu} F^2 \,.
\eeq
Taking the trace of the Einstein equation yields an on-shell expression for the regular part of the Ricci scalar
\beq
R_{bulk} = \frac{2D}{D-2}\Lambda + \frac{D-4}{D-2}  4 \pi G F^2\,.
\eeq
Hence, the bulk contribution to the on-shell action is  given by
\beq \label{bulkaction1}
I_{1/2,bulk} =-\frac{ 2}{   D-2}\int_{\mathcal{M}_{1/2}} d^D x \sqrt{g}\left (\frac{1}{8 \pi G} \Lambda -\frac{1}{4}F^2 \right)\,.
\eeq
Further, it can be shown that  the square of the electromagnetic field tensor \eqref{Fmunu} in   Euclidean static coordinates is
\begin{equation}
    F_{\mu\nu}F^{\mu\nu} 
    =- 2 \left ( \frac{Q}{   r^{D-2}} \right)^2\,.
\end{equation}
By inserting this into \eqref{bulkaction1} we find the bulk action  becomes  
\begin{equation}
   I_{1/2,bulk} = -\frac{\beta \Theta \Lambda}{8 \pi G (D-2)} - \frac{\beta Q }{2(D-2)} (\Phi_+   - \Phi_{++}  )\, .
\end{equation}
Here, we used equation \eqref{potentials} for the electric potentials, and we noted that the Euclidean spacetime volume $\int_{\mathcal M_{1/2}} d^D x \sqrt{g}$ is equal to $\beta \Theta/2$, where $\Theta$ is the Killing volume \eqref{eq:killingvolume}.

Second, integrating the  conical part of the Ricci scalar yields  \begin{equation}
    I_{1/2,con}= -\frac{\mathcal{A}_{+} \epsilon_{+}}{8 \pi G} - \frac{\mathcal{A}_{++}\epsilon_{++}}{8 \pi G} = - \frac{\mathcal{A}_{+}}{8G} - \frac{\mathcal{A}_{++}}{8G} + \frac{\beta}{2} \left ( \frac{\mathcal{A}_{+} \kappa_{+}}{8\pi G} + \frac{\mathcal{A}_{++} \kappa_{++}}{8\pi G} \right)\, , \end{equation}
where $\epsilon_{+,++}=\pi (1-n_{+,++})$ is the deficit angle  associated to the conical singularities. The range of the polar angle of one cap is $\pi n_{+,++} $, which is related to the Euclidean time period by $\pi n_{+,++} =\beta \kappa_{+,++}/2$. 
By adding the conical deficit and bulk terms  and using the Smarr formula \eqref{eq:rndssmarr}, the on-shell action becomes  
\begin{equation} \label{sumtwoactions}
    I_{1/2,bulk} + I_{1/2,con} = -  \frac{\mathcal{A}_+ }{8 G} - \frac{\mathcal{A}_{++}   }{8 G} -\frac{\beta Q}{2} (\Phi_+ - \Phi_{++})\, .
\end{equation}
Next let us add the boundary term of the Maxwell action, $I_{bdy}$, which originates from the fact that we are fixing the charge on the boundary, which is equivalent to fixing $n_\mu F^{\mu i}$ on the boundary \cite{Hawking:1995ap}. In order to compute the Maxwell boundary term we need to choose a gauge for the gauge potential. A suitable gauge choice that is regular everywhere on the Euclidean geometry is \cite{Mann:1995vb}
\begin{equation}
    A_1 = - i\frac{Q \tau}{r^{D-2}} dr  \,.
\end{equation}
Further, we choose coordinates on the boundary $\Sigma$ such that it is located at the surface $\tau =0,\beta/2$ and the coordinate $r$ runs from the black hole horizon to the cosmological horizon. Then, the boundary term can be computed to be
\begin{equation} \label{maxwellbdyaction}
    I_{1/2,bdy}=- \int_{\Sigma} d^{D-1} x \sqrt{h} F^{\mu\nu} n_\mu A_\nu=\frac{\beta Q}{2} (\Phi_+ - \Phi_{++})\,.
\end{equation}
Notice that the boundary term localizes on the two Euclidean horizons, and exactly cancels against the term proportional to $\beta$ in \eqref{sumtwoactions}.  
 Summing the three different contributions to the action  yields the total Euclidean action of $\mathcal M_{1/2}$
 \begin{equation}
     I_{1/2,RNdS}=  I_{1/2,bulk} + I_{1/2,con}+ I_{1/2,bdy}=-  \frac{\mathcal{A}_++\mathcal{A}_{++}}{8 G}\,.
 \end{equation}
 The action is the same for both halves of the Euclidean section, and hence the on-shell action of the entire Euclidean RNdS section $\mathcal M$  is given by minus the sum of the  black hole entropy and the cosmological horizon entropy
\begin{equation} 
\label{finalresultaction}
    I_{RNdS} =  -  \frac{\mathcal{A}_++\mathcal{A}_{++}}{4 G} = - S_{RNdS}\,.
\end{equation}
We emphasize that this expression is independent of the Euclidean time period $\beta$, and that it holds in any number of dimensions for arbitrary mass and charge. This is the action in the canonical ensemble where the temperature and charge are kept fixed.\footnote{In the grand canonical ensemble that fixes  the electric potential instead of the charge, there is no boundary term for the gauge field and the on-shell action of $ \mathcal M_{1/2}$ is given by \eqref{sumtwoactions} instead.} The same result for the on-shell action $I_{1/2,RNdS}$ was obtained in four dimensions   in \cite{Chao:1997osu,Wang:2022sbp}, which was used to compute the pair creation of fixed mass and charge black holes in de Sitter. 
We expect the creation of multiple charged black hole in de Sitter to be further suppressed with (a multiple power of) the same exponential factor, due to the fact that the exponential de Sitter expansion will create multiple independent static patches that contain just a single charged black hole. This expectation is in fact supported by the explicit construction of the (dynamical) solution for multiple de Sitter charged black holes \cite{Kastor:1992nn}, generalizing the Majumdar-Papapetrou solutions in flat space.
For zero charge the result \eqref{finalresultaction} is consistent with earlier findings for the action of Euclidean Schwarzschild-de Sitter space \cite{Chao:1997em,Bousso:1998na,Gregory:2014,SdSaction}. Further, the on-shell action of the extreme Euclidean solutions (proper instantons) was computed before in \cite{Mellor:1989wc,Hawking:1995ap,Mann:1995vb,Cardoso:2004uz}, which we discuss in more detail below.


\subsection{Extreme cases:  cold,   lukewarm, Nariai and ultracold instantons}

In the literature the    action is usually only evaluated for the extreme Euclidean RNdS solutions:   cold, lukewarm, Nariai and ultracold   instantons. This is because they are proper stationary points of the Euclidean Einstein action, i.e., they are regular  and hence satisfy the Einstein equation  everywhere. Below we specialize to these true instanton cases and compare our results for the on-shell action to those in the literature, especially   the  comprehensive paper by Mann and Ross \cite{Mann:1995vb}.  
From our general expression \eqref{finalresultaction} it follows that the on-shell actions of the  extremal, lukewarm, Nariai and ultracold instantons  are,  
\begin{align}
I_E&= - \frac{\mathcal{A}_{E,+}+\mathcal{A}_{E,++}}{4G} = - \frac{\Omega_{D-2}}{4G}\,\left (r_E^{D-2}+r_{E,++}^{D-2}\right)\,, \label{extremallaction}\\
 I_L&= - \frac{\mathcal{A}_{L,+}+\mathcal{A}_{L,{++}}}{4G} = - \frac{\Omega_{D-2}}{4G}\,\left(r_{L,+}^{D-2}+r_{L,{++}}^{D-2}\right)\,, \label{lukewarmaction}\\
 I_N&= - 2 \frac{\mathcal{A}_{N}}{4G} = - \frac{\Omega_{D-2}}{2G}\,r_N^{D-2}\,, \label{Nariaiaction}\\
 I_U &= - 2 \frac{\mathcal{A}_{U}}{4G}=- \frac{\Omega_{D-2}}{2G}\,r_U^{D-2} \label{ultracolldaction}\,.
\end{align}
Obviously, since  the expression \eqref{finalresultaction} for the Euclidean action is a continuous function of the mass and charge, the action of the proper instantons is also equal to minus the sum of the black hole and cosmological horizon entropy. On the other hand,  Mann and Ross computed the action of each proper instanton separately by using the metric of each instanton.  Their results for the Euclidean action of the lukewarm and charged Nariai   instantons agree with only half  the expressions above, in particular they agree with     \eqref{lukewarmaction} and \eqref{Nariaiaction}, but not with \eqref{extremallaction} and \eqref{ultracolldaction}.   Furthermore,  following earlier work by Hawking, Horowitz and Ross \cite{Hawking:1994ii}, they found for the extremal instanton that the   black hole horizon makes no contribution to the action and hence to the entropy. This is because   only the cosmological horizon is part of the instanton, and the black hole horizon is infinitely far away from any other point.  Thus, the extremal black hole entropy vanishes according to them, even though its horizon area   is nonzero (see also \cite{Carroll:2009maa}). This is in disagreement, however,  with the modern view that the  entropy of extremal black holes is finite in the classical regime (see, e.g., \cite{Nayak:2018qej}).\footnote{Quantum corrections, in fact, flip this story again, since recent developments (see, e.g., \cite{Turiaci:2023wrh}) show that for non-supersymmetric Reissner-Nordstr\"{o}m black holes the density of states   (hence, the black hole entropy) at fixed charge   becomes exceedingly small near extremality.} Moreover, a zero extremal black hole entropy would imply that the RNdS action is discontinuous as a function of the mass and charge, which contradicts our result for the action. The same confusion about extremal black hole entropy seems to arise in the more recent papers \cite{Dias:2004px,Wang:2022sbp} on the  Euclidean RNdS action.

 \begin{figure}[t]
  \centering
  \includegraphics[width=8.0cm]{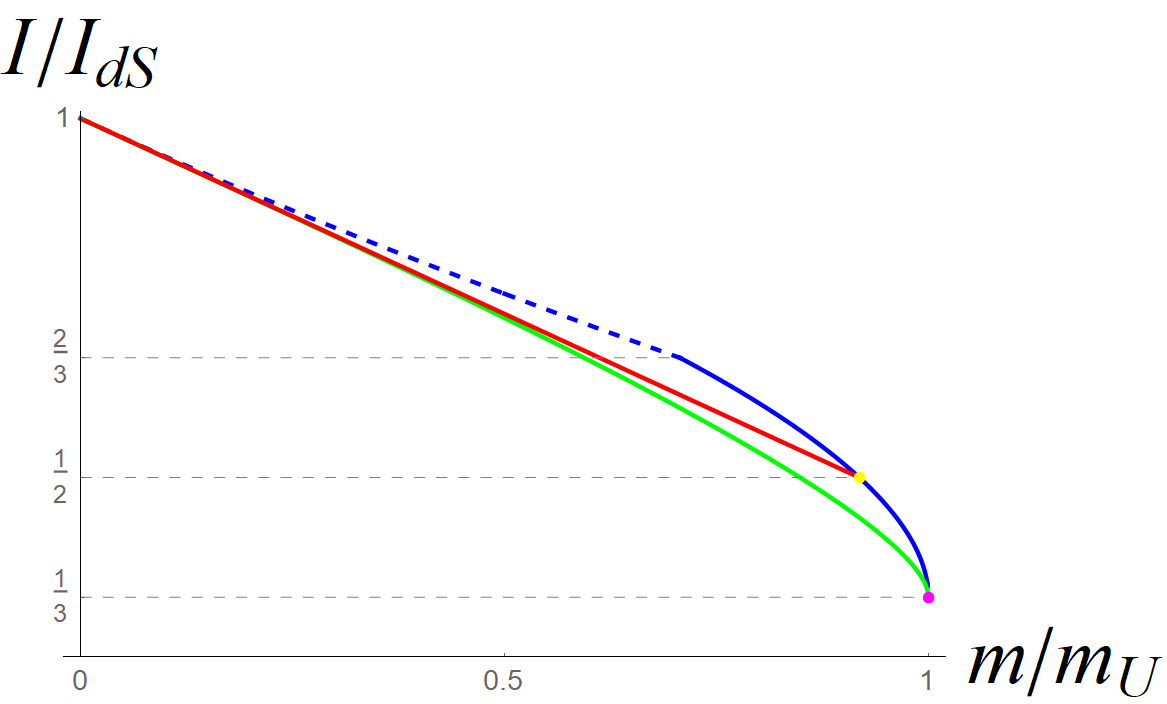}
  \caption{Fraction of the Euclidean action of proper instantons over the action of pure de Sitter  $I/I_{dS}$ vs. the mass parameter $  m/ m_U$. The blue line corresponds to charged Nariai $I_N$, the green line to the extremal solution $I_E$, and the red line to the lukewarm solution $I_L$. The blue dashed curve corresponds to the action of the Schwarzschild-de Sitter black hole. The special points $I_{LN}$ and $I_{U}$ are shown as yellow and magenta dots, respectively.}
  \label{shark2}
\end{figure}

Furthermore,  for the ultracold instanton Mann and Ross obtained two different results for the on-shell action  depending on the near-horizon metric. The action of one of the ultracold metrics, called $UC1,$ agrees with the limit of their action of the extremal solution as it approaches the ultracold solution, but it is not the same as the ultracold limit of the action of the Nariai solution. The   action of the other ultracold metric $UC2$  vanishes identically, since the   metric describes $2D$ Minkowski spacetime (times a sphere) in standard coordinates without  horizons. Our result does not agree with either of these  actions. In fact, our action for the ultracold solution can be obtained as a special case of the action  for both the extremal solution and the charged Nariai solution. In Figure \ref{shark2} we plot the fraction of the   action of the proper instantons and the action for de Sitter space against the dimensionless mass $m/m_U.$ One can see that the  ultracold point (magenta point) is connected to both the Nariai branch (blue line) and the extremal branch (green line). This plot should be contrasted with Figure 3 in the paper by Mann and Ross \cite{Mann:1995vb}, in which  the Nariai branch and extremal branch are not connected.

\begin{figure}[t]
  \centering
  \includegraphics[width=7.5cm]{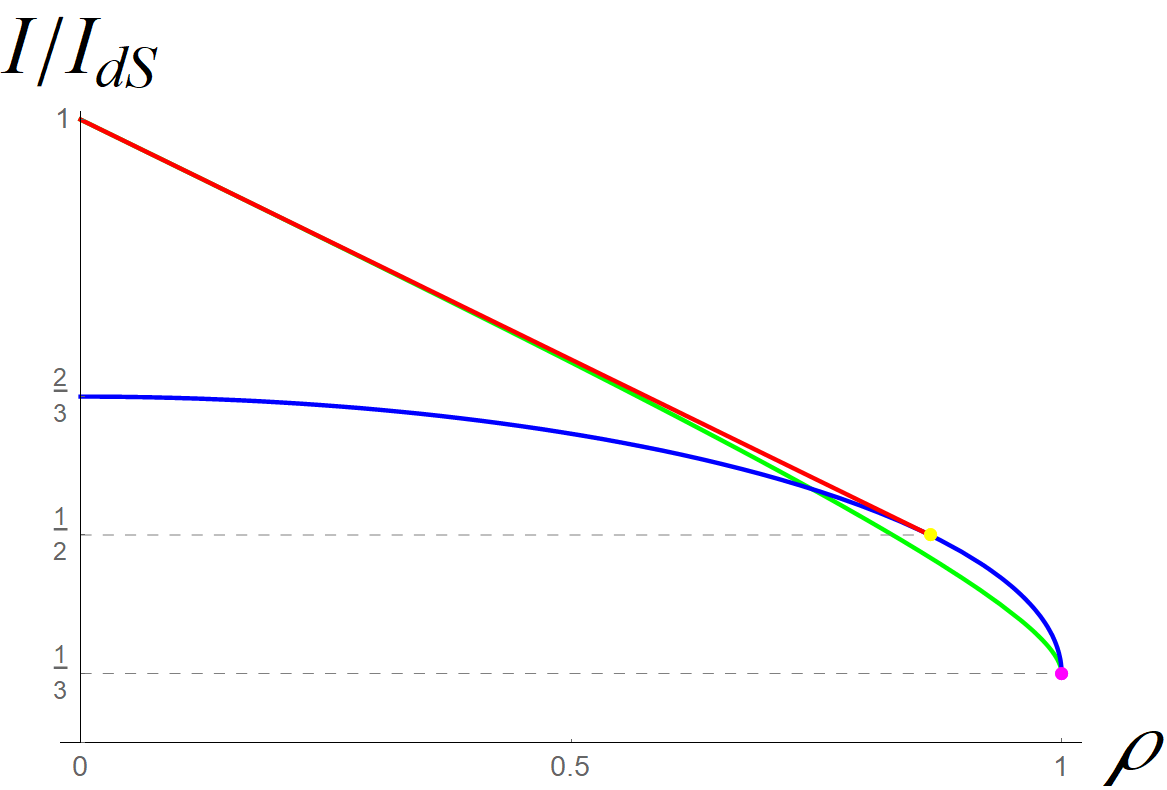}
  \caption{Fraction of the Euclidean action of proper instantons over the action of pure de Sitter  $I/I_{dS}$ vs. the charge parameter $\rho$. The blue line corresponds to charged Nariai~$I_N$, the green line to the extremal-cold solution $I_E$, and the red line to the lukewarm solution~$I_L$. The special points $I_{LN}$ and $I_{U}$ are shown as yellow and magenta dots, respectively.}
  \label{sharkplac}
\end{figure}

As a function of the mass the action of the proper instantons is   cumbersome. But in terms of the charge   the Euclidean on-shell action   takes a relatively simple form  in $D=4$  
 \begin{align}
   I_E &=- \frac{\pi}{\Lambda G} \left ( 3 - \sqrt{2 + 2 \rho_E^2 - 2 \sqrt{1 - \rho_E^2}}\right) \,,\\
   I_L &=- \frac{\pi}{  \Lambda G} \left ( 3 -   \sqrt{3}   \rho_L \right) \,,\\
  I_N &= - \frac{\pi}{\Lambda G} \left (1 + \sqrt{1 - \rho_N^2} \right) \,,\\
 I_{dS} &= - \frac{3 \pi}{\Lambda G}\,, \quad I_{NN} = - \frac{2\pi}{\Lambda G}\,,  \quad I_{LN} =- \frac{3 \pi}{2 \Lambda G}\,, \quad   I_U  = -\frac{\pi}{\Lambda G}\,,
\end{align}
where $I_{NN}$ denotes the action of the neutral Nariai solution ($\rho_{NN}=0$) and $I_{LN}$ the action of the lukewarm-Nariai solution ($\rho_{LN}=\frac{\sqrt{3}}{2}$).  In   Figure~\ref{sharkplac} we plot the fraction of the proper instanton actions and the de Sitter action against the charge parameter $\rho =q/q_U$.   We note the actions of the lukewarm (red line) and cold (green line) solutions coincide at $\rho_{dS}=0$ and are equal to the de Sitter action $I_{dS}$ in that case. Further, the actions of the lukewarm and Nariai (blue line) solutions coincide   for the lukewarm-Nariai solution (yellow point). Moreover, we can   see from the figure that the actions of the extremal and Nariai solutions are the same at $\rho = 1$, which corresponds to the ultracold solution (magenta point). The extremal and Nariai solutions also  have the same action at $\rho = \frac{\sqrt{5}}{3} $, but this is not a  special point in the shark fin diagram. Figure~\ref{sharkplac} can be contrasted with Figure 3 in a paper by Bousso and Hawking \cite{Bousso:1996au} where the fraction $I/I_{dS}$ is plotted against the square of the charge parameter (in fact, the dimensionless quantity $q^2 \Lambda).$ In that figure   by Bousso and Hawking (which is based on the computations of the action   by Mann and Ross \cite{Mann:1995vb}) the Nariai branch and the extremal branch do not meet at the ultracold point, which   seems at odds with the shark fin diagram. 
 
\section{Constrained instantons and the pair creation rate \label{secCons}}

Having computed the on-shell Euclidean action for charged black holes in de Sitter space in any number of dimensions, it is clear that, for general charge and mass, they can only be understood as proper stationary  instanton  solutions after introducing a constraint. 
After a very brief introduction of the general method of constrained instantons \cite{AFFLECK1981429,Cotler:2020lxj,Cotler:2021}, we will proceed in the same way as in our previous work \cite{SdSaction} to compute the pair creation rate of charged black holes in de Sitter using a constrained path integral. Except that now, in addition to the introduction of a functional that implements fixed mass, we also have to introduce a functional that fixes the charge. In essence, as also emphasized by the construction of the Hartle-Hawking wavefunction of RNdS, the presence of conical singularities implies ``conical boundary'' contributions to the on-shell Euclidean action where the fields are not allowed to vary, translating to fixed mass and charge, and necessary in  order to interpret the solution as a  stationary contribution to the partition function. 

To understand the contribution of the constrained stationary solutions to the Euclidean gravitational partition function, we introduce the mass and charge constraint in terms of a delta function and integrate over them to rewrite the partition function as follows 
\begin{equation}
    \int[dg][dA] \, e^{-I[g,A]} = \int d\alpha \int d\rho \int [dg][dA] \, \delta({\cal C}[g,A]-\alpha) \, \delta({\cal D}[g,A]-\rho) e^{-I[g]} \, .
\end{equation}
Here $g$ and $A$ refer to the metric and gauge field degrees of freedom and $I$ is the Euclidean off-shell action. Although the above expression is completely general, what we have in mind is that the constraint functional ${\cal C}[g,A]$ fixes the mass to be $m$, whereas the constraint functional ${\cal D}[g,A]$ fixes the charge to be $q$. Here, as before, the dimensionless ratio $\alpha$ runs from $0$ to $1$ for any charge $q$, and $\rho$ equals $q/q_U$, where $q_U$ corresponds to the maximum ultracold charge. The next step is to introduce the integral representation of the delta functions to impose the constraints by means of two Lagrange multiplier terms added to the action
\begin{equation} 
\int[dg][dA] \, e^{-I[g,A]} = \int d\alpha \int d\rho \int d\lambda_m \int d\lambda_q \int [dg][dA] \, e^{-I[g,A]+\lambda_m({\cal C}[g,A]-\alpha) + \lambda_q({\cal D}[g,A]-\rho))} \, ,
\end{equation}
where it should be understood that the integral over the Lagrange multipliers $\lambda_m$ and $\lambda_q$ are parallel to the imaginary axis. With this formal rewriting of the general partition function we can now identify the contributions from the charged de Sitter black holes in terms of an actual saddle point approximation. For any fixed $\alpha$ and $\rho$, only allowing $\lambda_m$, $\lambda_q$ and the degrees of freedom $g$ and $A$ to vary, the stationary points of the new action, including the Lagrange multiplier terms, correspond to solutions of the following (schematic) equations
\begin{eqnarray}
    &&\delta_g I[g,A] + \lambda_m \, \delta_g {\cal C}[g,A] + \lambda_q \, \delta_g {\cal D}[g,A] = 0 \, , \nonumber \\
    &&\delta_A I[g,A] + \lambda_m \, \delta_A {\cal C}[g,A] + \lambda_q \, \delta_A {\cal D}[g,A] = 0 \, , \nonumber \\
    &&{\cal C}[g,A]=\alpha \, , \quad {\cal D}[g,A]=\rho \, . 
\end{eqnarray}
As in the neutral SdS case, the important point is that the charged black hole is only a solution for non-vanishing $\lambda_m$ and $\lambda_q$, where both Lagrange multipliers compensate for variations of the mass and charge respectively, in such a way that the solution is now a true stationary point of the modified action. That then allows us to use the saddle point approximation, valid at weak gravitational coupling, to estimate the partition function by integrating over all the saddles parameterized by the two constraint parameters, the mass and the charge respectively. The constrained instanton interpretation for black holes in de Sitter space was already anticipated in \cite{Chao:1997em, Chao:1997osu, Bousso:1998vz}, as well as more recently in \cite{Draper:2022xzl}, and after   applying it to the neutral case in \cite{SdSaction}, here we are pointing out that it can straightforwardly be generalized to charged black holes in de Sitter space as well. Note that the evaluation of the new action, featuring the constraints, will simply reproduce the on-shell Euclidean action of the charged black hole that we computed in the previous section. 

Importantly, any set of constraint functionals effectively fixing the mass and charge should give the same result for the partition function. Since we will only be interested in the exponential dependence on mass and charge as given by the horizon entropies, and ignore any contributions coming from the one-loop determinant, an explicit construction of the constraints will not be necessary. We will assume that constraint functionals linear in the mass and charge can be constructed, meaning that we can integrate over the dimensionless variables $\mu$ and $\rho$ to determine the contributions to the partition function without any additional dependence on the mass and charge coming from the Gaussian path integral (the one-loop determinant). 

\begin{figure}[t]
\centering
\begin{subfigure}{.5\textwidth}
  \centering
  \includegraphics[width=6.5cm]{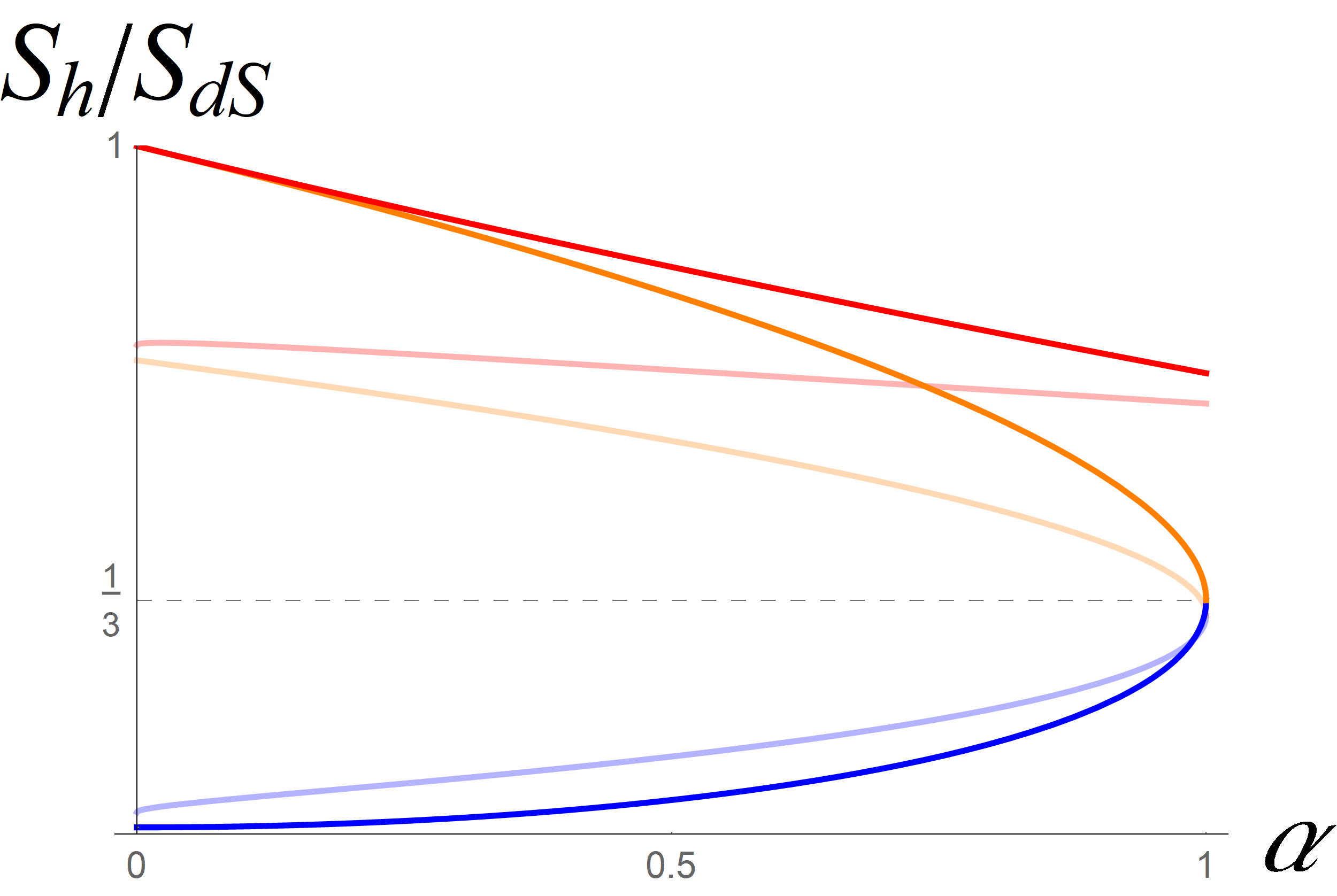}
  \caption*{a) $\rho=0,1/2$.}
\end{subfigure}%
\begin{subfigure}{.5\textwidth}
  \centering
\includegraphics[width=6.5cm]{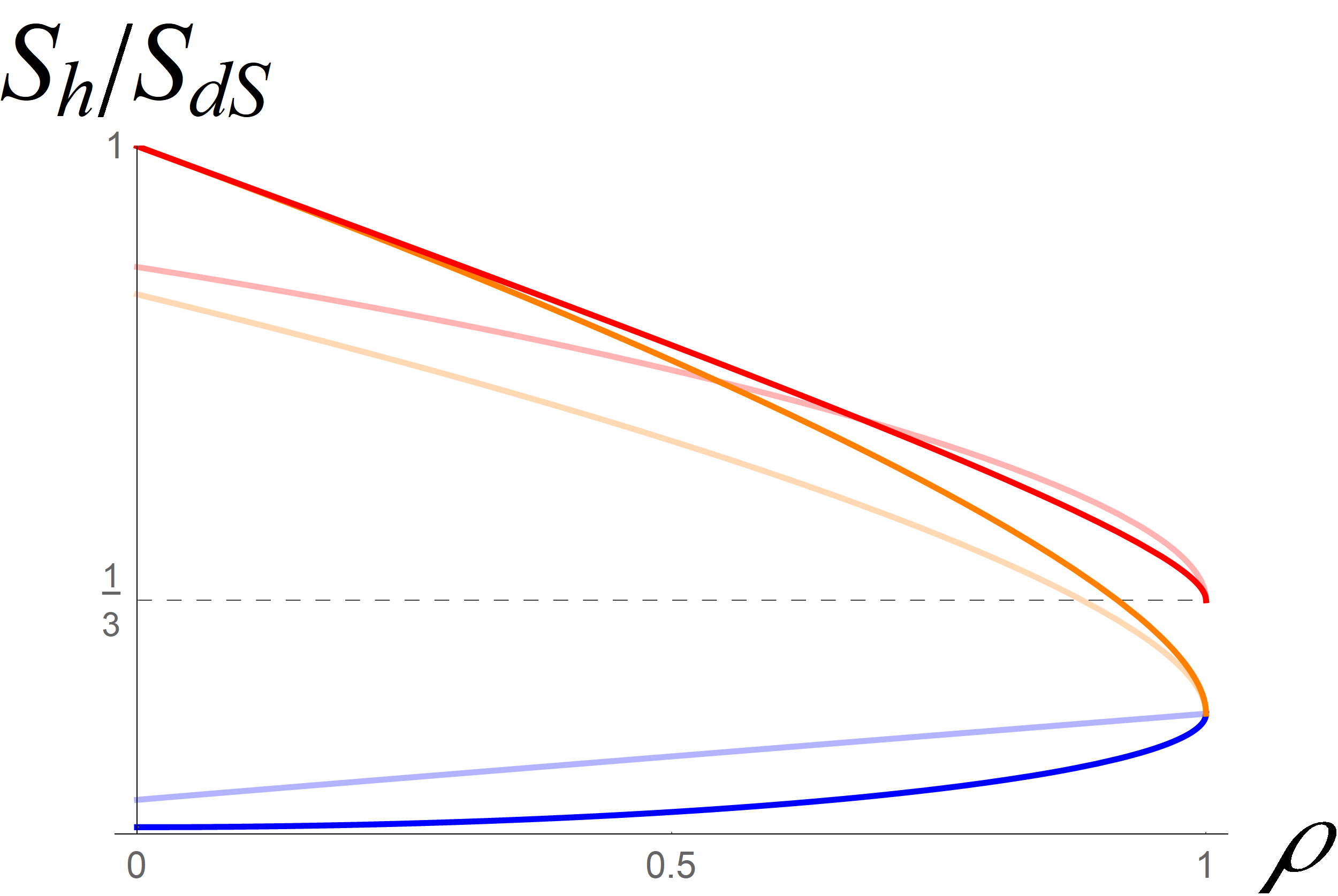}
  \caption*{b) $\alpha=0,1/2$}
\end{subfigure}
\caption{Horizon entropies (a) as a function of  the mass parameter $\alpha$ for fixed $\rho=0,1/2$, and (b) as a function of    the charge parameter $\rho$   for $\alpha=0,1/2$. The black hole entropy $S_+$ is shown in blue, the cosmological horizon entropy $S_{++}$ in orange, and the total entropy $S_{RNdS}=S_++S_{++}$  in red. Dark  colors correspond to $\alpha,\rho=0$, and light  colors to $\alpha,\rho=1/2.$}
\label{fig:entroshades}
\end{figure}

We are now ready to use this saddle point approximation of the gravitational partition function to derive an expression for the probability to spontaneously nucleate charged black hole of arbitrary mass and charge in de Sitter space. By integrating over the charge and mass, and comparing to the on-shell action of empty de Sitter space, we get 
\begin{equation}
    \Gamma=\int d\alpha\,d\rho \, e^{-(I_{RNdS}-I_{dS})} =\int d\alpha\,d\rho\, e^{-(S_{dS}-S_{RNdS})} \,.
\end{equation}
Unfortunately, as in the neutral case \cite{SdSaction}, it is not possible to analytically perform this integral without making some approximations. From Figure \ref{fig:entroshades}, as in the neutral case, one notices that the entropy is roughly linear in $\alpha$, at a fixed value for the charge. So  we will fit the total entropy $S_{RNdS}$ as a linear function of $\alpha$, at fixed charge ratio $\rho$, as follows
\begin{align}
    S_{fit} &\equiv \alpha \,S_N+(1-\alpha)\,S_E
    \, , 
\end{align}
where the extremal entropies $S_N$ and $S_E$ are given by   
\begin{equation}
    S_{E,N} (\alpha_{E,N}, \rho)\equiv\frac{\mathcal{A}_{+}(\alpha_{E,N}, \rho)+\mathcal{A}_{++}(\alpha_{E,N}, \rho)}{4G} \, .
\end{equation}
This linear approximation, for fixed charge, gets better as the number of dimensions increases (at worst, in $D=4$, the relative difference is $1.4 \%$ at $\alpha=0.23$ and $\rho=0.81$).
 
Armed with this linear fit we can now straightforwardly perform the integral over $\alpha$. The transition rate from empty de Sitter space to an arbitrary mass black hole, in a fixed charge sector labeled by $\rho$, becomes
\begin{align}
    \Gamma_{\rho}&\approx\int_{0}^{1}\,d\alpha\,\,e^{-(S_{dS}-S_{fit})}
    = e^{-(S_{dS}-S_E)}\left(\frac{1-e^{S_N-S_E}}{S_E-S_N}\right)\, .
\end{align}
Obviously, this reduces to our previous result \cite{SdSaction} for $\rho=0$ as $S_E(\rho=0)=S_{dS}$. In the opposite limit $\rho=1$, the extremal entropies $S_E$ and $S_N$ become equal to $S_U=\mathcal{A}_U/(2G)$ and the result is proportional to the entropy deficit between the ultracold solution and empty de Sitter space, as one might have expected. Now, to compute the creation rate of arbitrary massive \textit{and} charged black holes, one would be left with computing the following integral 
\begin{align}
    \Gamma&=\int_{0}^{1}\,d\rho\, \Gamma_{\rho}\,,
\end{align}
for which it is challenging to find an expression in closed form. Introducing, however, another linear approximation, this time for the extremal entropies $S_{E,N}$ as a function of the charge ratio $\rho$
\begin{equation}
    S_E\approx \rho\, S_U + (1-\rho) S_{dS} \, , \qquad S_N\approx  \rho\, S_U + (1-\rho) \, S_{N}^0\, ,
\end{equation}
where $S_U=\mathcal{A}_U/(2G)$ is the ultracold entropy and $S_N^0=S_N(\rho=0)$ the entropy of the neutral Nariai black hole, a two-dimensional linear fit $\Tilde{S}_{fit}$ for the entropy takes the following form
\begin{equation}
    \Tilde{S}_{fit}=\rho\, S_U+\alpha\,(1-\rho)\,S_N^0+(1-\alpha)(1-\rho)\,S_{dS}\, .
\end{equation}
Note that this linear fit is not a good approximation everywhere in the parameter space of the charged de Sitter black hole. In fact, this fit for the entropy is clearly worse as compared to the one for fixed charge (in $D=4$, the maximal relative difference is now $13.6\%$ at $\alpha=1$ and $\rho=0.77$). Even so, we believe it is useful to allow for an approximation of the creation rate in closed form. Indeed, using this fit we can now analytically compute the integral and obtain
\begin{align}
\label{eq:totcrea}
  \Gamma\approx \frac{e^{-(S_{dS}-S_{U})}}{S_{dS}-S_N^0}\bigg(&\gamma(\sigma_N)-\gamma(\sigma_{dS}) +\log{(\sigma_N/\sigma_{dS})}\bigg) \, ,
\end{align}
where $\sigma_{N}\equiv S_U-S_N^0$ and $\sigma_{dS}\equiv S_U-S_{dS}$, and $\gamma(\sigma)$ corresponds to the incomplete Euler function $\gamma(a=0,\sigma)$, that we relabeled in order to not be confused with the creation rate
\begin{equation*}
    \gamma(\sigma)\equiv \int_{\sigma}^{\infty} t^{-1} e^{-t} \, dt\,.
\end{equation*}
The expression (\ref{eq:totcrea}) gives a reasonable first estimate in closed form for the total creation rate.   However, we should stress that these expressions for the creation rates are merely indicative, in the sense that they could receive several corrections. First of all, we expect corrections to the linear approximations we used. Secondly, we assumed for simplicity the Gaussian integral over the stationary solutions does not introduce additional mass or charge dependence. However, we expect that the one-loop determinant does depend on the mass and charge, which would alter our closed form for the  pair creation rate. The latter is clearly important to check in future work. 

\section{A comment on the Festina Lente bound \label{wgc}}

\begin{figure}[t]
  \centering
  \includegraphics[width=9cm]{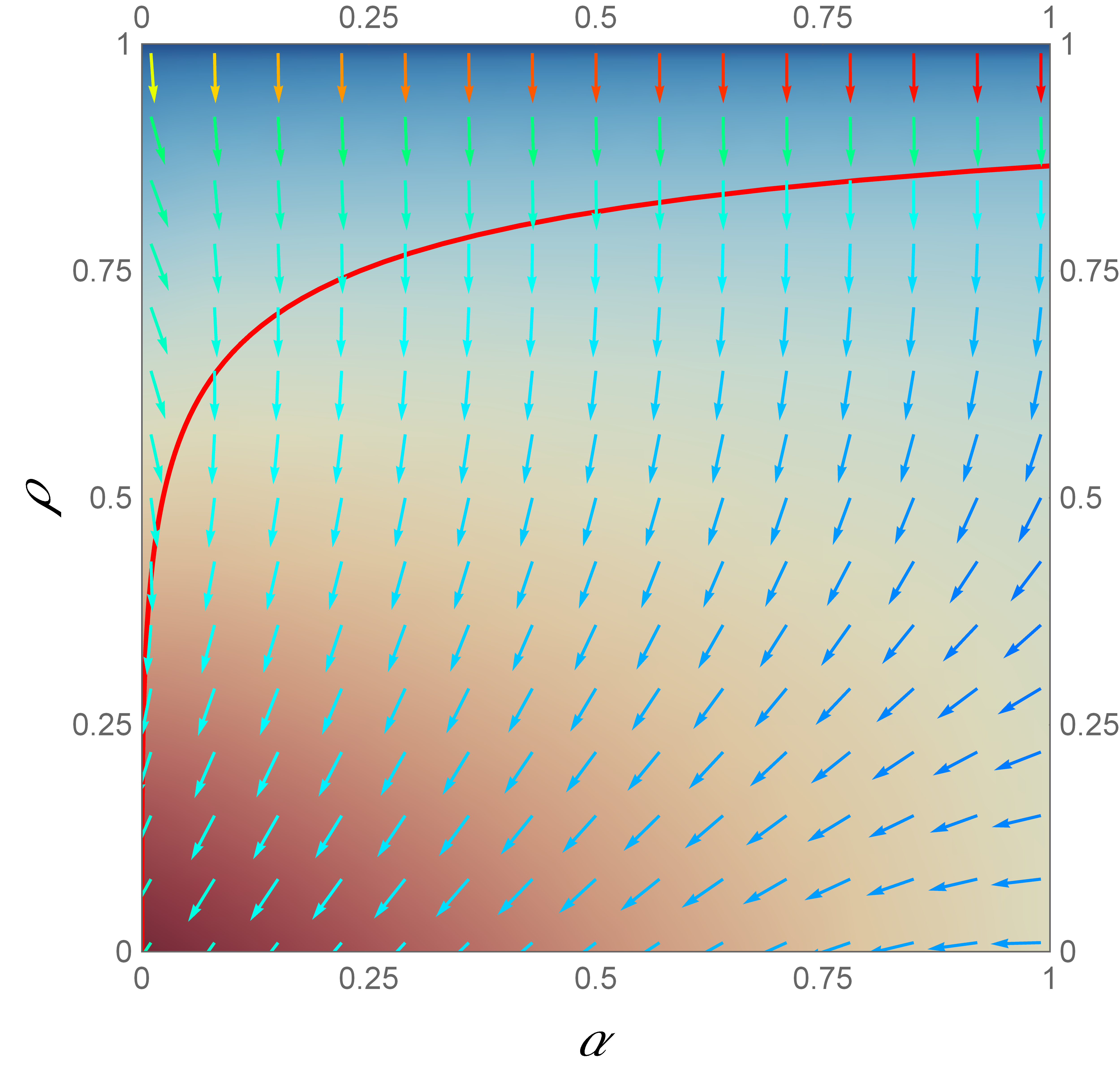}
  \caption{RNdS entropy $S_{RNdS}$ in terms of $\alpha$ and $\rho$, where reddish colors correspond to high values, whereas blueish colors to low values. The arrows stand for the gradient and indicate a   tendency to escape from the ultracold region at the top. Notice, that the lukewarm path (red line) separates the arrows pointing left from those pointing right (losing or gaining mass, respectively).}
  \label{phasentro}
\end{figure}

In flat space, as well as Anti-de Sitter space, the Weak Gravity Conjecture is based on the idea that any black hole should be able to decay, even if the black hole is extremal and has a vanishing temperature, without producing a remnant naked singularity, in line with the Cosmic Censorship Conjecture (see \cite{Wald:CC} and references therein). This puts a lower bound on the charge $q_e$ to mass $m_e$ ratio of an elementary excitation, explaining why the gravitational force is weak relative to the electromagnetic force. 

In de Sitter space the configuration space of charged black holes has a more elaborate structure (see Figure~\ref{shark}), implying that the Cosmic Censorship Conjecture should forbid both decay by emitting too much charge $\delta q$, as well as emitting too much mass $\delta m$, to avoid ending up in the region to the right of the Nariai boundary or to the left of the extremal boundary, respectively. Moreover, since empty de Sitter corresponds to the maximum entropy state, one might expect the effective black hole decay to be described by thermodynamics, i.e., the gradient of the entropy function (see Figure~\ref{phasentro}). This suggests that the probability of decay from a macroscopic black hole with mass and charge parameters $(m,q)$ to one with parameters $(m',q')=(m-\delta m, q-\delta q)$ is always proportional to $\exp[\Delta S]$, with $\Delta S \equiv S(m',q')-S(m,q)$. 

As explained in \cite{Aalsma:2018}, using an s-wave tunneling formalism in flat and AdS space, this is what one indeed finds, even in the so-called ``superradiant regime'' at low energies, where $\Delta S>0$, describing a fast, entropically favourable, decay of the charge of the black hole. We expect an analogous line of reasoning to hold just as well in de Sitter space. Indeed, in \cite{Montero:2019ekk} one concludes, supported by explicit calculations of the Schwinger pair creation rate, that the adiabatic, small mass regime should be avoided,\footnote{This should coincide with the superradiant regime, as explained in \cite{Aalsma:2018}. 
As for rotating black holes, the superradiant regime for charged black holes allows energy to be extracted by scattering a beam of charged particles off the charged black hole, through stimulated pair creation in a region close to the black hole. This charged black hole `generalized ergosphere' region will depend on the properties of the charged particle flux, as opposed to the universal ergosphere of a rotating black hole \cite{Denardo:1973pyo}.} 
otherwise the decay to singular states to the right of the Nariai boundary would be possible, violating the Cosmic Censorship Conjecture and prohibiting an effective description of the (slow) decay. This sets a lower ``Festina Lente'' bound on the mass of an elementary excitation, in contrast to the Weak Gravity Conjecture, which sets an upper bound on the mass of an elementary excitation.

This supports the conjecture that an s-wave tunneling calculation in de Sitter space should be able to reproduce the effective evolution inside the allowed shark fin region, as long as one excludes the superradiant regime by requiring the absence of charged particles that are too light. In other words, we would like to suggest that the Festina Lente bound can be derived (and understood) more precisely by identifying the microscopic mass and charge parameters for which $\Delta S$ vanishes. This would be interesting to check explicitly, especially in the ultracold limit, where the macroscopic effective decay, described in terms of $dq/dm$, is fixed uniquely and equals  
\begin{align}
\label{eq:wgcultra}
  \left. \frac{dq}{dm} \right|_{U} =\frac{\sqrt{D-2}}{2} \, .
\end{align}
It is important to realize that this unique limiting direction of the effective decay in the top region of the shark fin should be obtained for a range of elementary charges and masses, as long as a superradiant and sub-extremal regime in the mass and charge parameter space is excluded. We suspect that a tunneling calculation can allow for a precise determination (including backreaction) of the superradiant regime and therefore a more precise (and better understood) version of the Festina Lente bound (in arbitrary dimensions). 

\section{Summary and conclusions}
\label{sec:conclusion}

Let us summarize our results. First of all, using exactly the same general techniques as for the neutral, static de Sitter black hole, we have shown that the on-shell Euclidean action of charged, static de Sitter black holes in arbitrary dimensions equals minus the sum of the black hole and the cosmological horizon entropy. This expected result relies on an important subtlety. Namely that the presence of conical singularities implies the necessary inclusion of (gauge field) boundary terms, providing a localized contribution to the on-shell action on the conical singularities that precisely cancels the contribution to the action from the bulk term depending on the charge and the gauge potentials.  As in the neutral case, this again emphasizes the intriguing role of conical singularities in Euclidean quantum gravity. This general expression, valid for any charge and mass, also helps to clarify some confusion in the literature \cite{Mann:1995vb,Cai:1997ih,Dias:2004px,Wang:2022sbp} for the extremal and ultracold black hole Euclidean actions in de Sitter, which can now be understood as a limit of the general expression, with a clear and unambiguous answer. 

Furthermore, equipped with the on-shell Euclidean action and an understanding in terms of constrained instantons, we used a saddle point approximation to compute an analytic expression for the probability to spontaneously nucleate an arbitrary mass and charge black hole in de Sitter space. In order to be able to compute this we used a linear approximation in the mass and charge of the entropy deficit. Although this is certainly not a good approximation everywhere in the full configuration space of charged de Sitter black holes, it does provide a reasonable result in closed form. We should stress that we expect corrections, in particular we ignored any potential contributions from the one-loop determinants around the two-parameter family of saddle-points. We hope to come back to this point in the future, starting with the neutral case. 

Finally, with the result for the Euclidean action, in arbitrary dimensions, and the expectation that thermodynamics should be governing the decay of charged black holes in de Sitter, we argued that a tunneling calculation similar to \cite{Aalsma:2018} should be able to explicitly reproduce the slow effective decay of charged de Sitter black holes within the allowed shark fin region as long as one avoids the superradiant regime. This would be particularly interesting to check in the ultracold limit, where the effective evolution describing the decay is fixed uniquely. We conjecture that by using the tunneling formalism it should be possible to derive a more precise version of the Festina Lente type bound in arbitrary dimensions (see our follow-up paper \cite{Aalsma:2023mkz}). 

Last, but not least, this study has further explored the space of non-perturbative corrections in de Sitter space, as described by saddle points in a Euclidean (quantum) gravity description. Although AdS/CFT has provided strong hints that saddle points in the partition function of Euclidean Einstein gravity have an important role to play, it is unclear whether this can be generalized to de Sitter space. Nevertheless, we believe it is important to further study the potential implications of these non-perturbative contributions, in particular how they might contribute to relevant observables, such as late-time correlators~\cite{Aalsma:2022eru}. The hope is, of course, that these corrections are such that these observables behave as they are expected from a unitary (holographic) microscopic perspective. 

\acknowledgments
We thank Lars Aalsma, Miguel Montero, Thomas van Riet and Gerben Venken for useful discussions and  correspondence. This work is part of the Delta ITP consortium, a program of the Netherlands Organisation for Scientific Research (NWO) that is funded by the Dutch Ministry of Education, Culture and Science (OCW). EM is supported by Ama Mundu Technologies (Adoro te Devote Grant 2019).
MRV is supported by  SNF Postdoc Mobility grant P500PT-206877 ``Semi-classical thermodynamics of black holes and the information paradox''.

\appendix

\section{Four dimensions}
\label{4D}

In this appendix we specialize to four dimensions and provide some potentially useful expressions for the horizon radii, mass, charge and temperatures of the RNdS solution.

\subsection*{Horizon radii}
In four dimensions the blackening factor can be written as
\begin{equation}
    f(r)=\frac{(r-r_{+})(r-r_{++})(r-r_{-})(r-r_{--})}{l^2\,r^2}\, .
\end{equation}
The outer  horizon radius $r_{+}$ and cosmological horizon radius $r_{++}$ can be expressed as
\begin{equation}
\label{chargedradii}
 r_{+,++}=r_{U}\left(x\mp\sqrt{3-x^2-2 m/(x\,m_{U})}\right)\, ,
\end{equation}
where $x\equiv\sqrt{1+\sqrt{1-\rho^2}\cos{2\eta}}$ with  $\eta\equiv\frac{1}{3}\arccos{\sqrt{(m_E^2-m^2)/(m_E^2-m_N^2)}}$. 
\begin{figure}[t]
\centering
\begin{subfigure}{.5\textwidth}
  \centering
  \includegraphics[width=6.5cm]{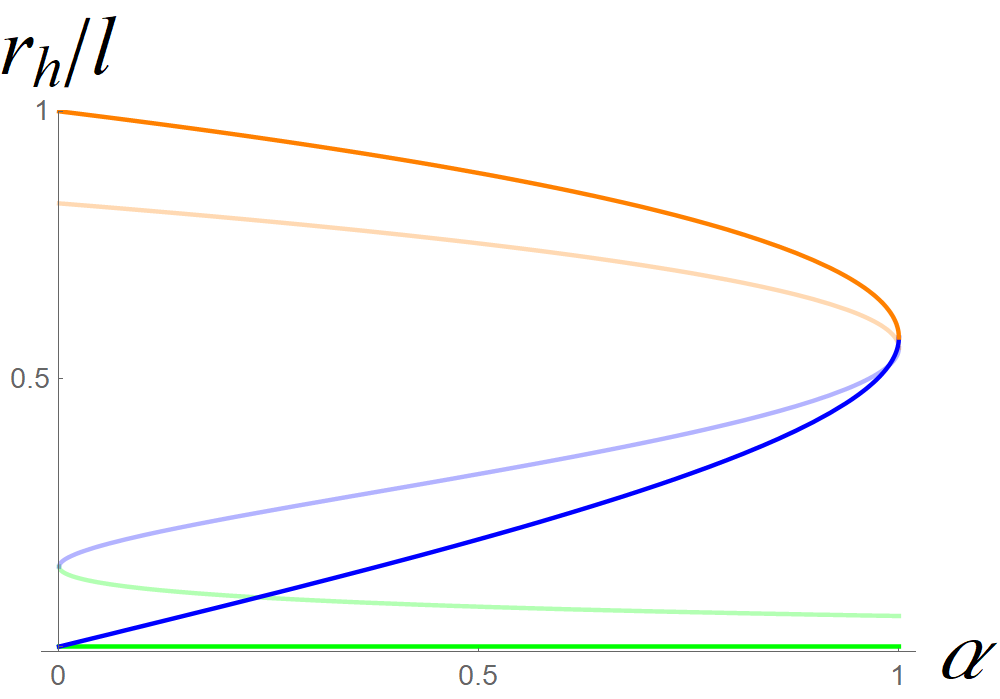}
  \caption*{a) $\rho=0,1/2$.}
  \label{fig:c0}
\end{subfigure}%
\begin{subfigure}{.5\textwidth}
  \centering
  \includegraphics[width=6.5cm]{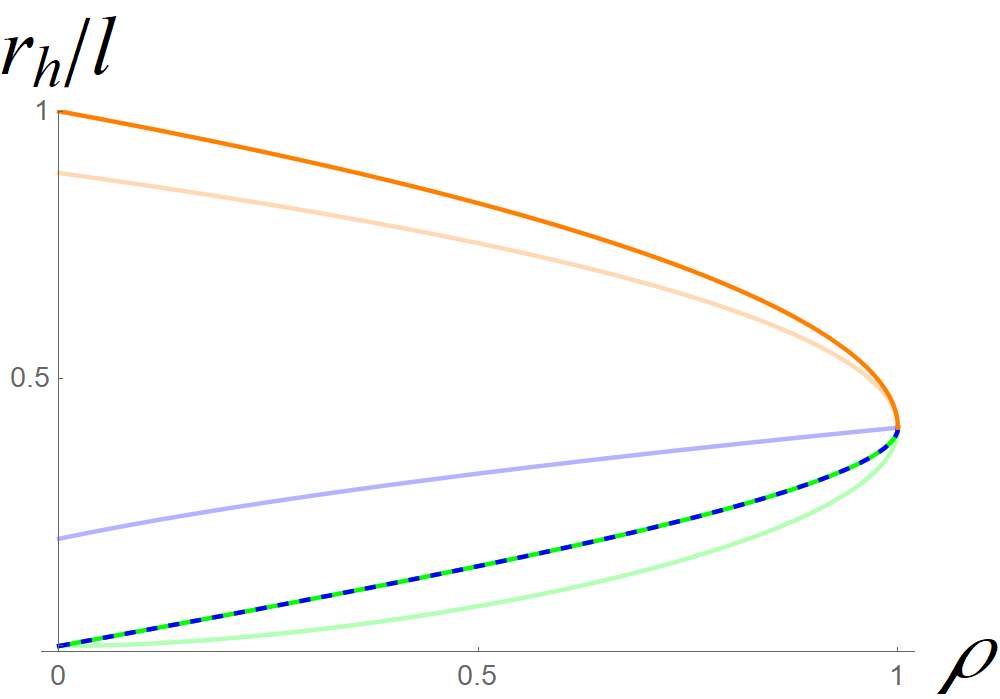}
  \caption*{b) $\alpha=0,1/2$}
  \label{fig:c736}
\end{subfigure}
\caption{Horizon radii (a) as a function of the mass parameter $\alpha$ and (b) as a function of the charge parameter $\rho$. The inner horizon radius $r_{-}$ is shown in green, the outer horizon radius $r_+$ in blue, and  the cosmological horizon radius $r_{++}$ in orange.  Dark  colors correspond to $\alpha,\rho=0$, and light  colors to $\alpha,\rho=1/2.$
Note that for $\alpha=0$, the inner and outer horizons are identical (dashed on the right).} 
\label{fig:radiishades}
\end{figure}
Moreover, the horizon radius, mass and charge of the ultracold black hole  are given by
\beq
r_U=\frac{l}{\sqrt{6}}\,, \qquad m_U=l\,\left(\frac{2}{3}\right)^{\frac{3}{2}}\,, \qquad q_U=\frac{l}{2\sqrt{3}}\,.
\eeq 
The  mass   for the   extremal and Nariai cases   as a function of the charge $\rho$ is  
\begin{align}
\label{eq:men}
    m_{E,N}=
    2 r_U\sqrt{1+3\rho^2\mp(1-\rho^2)^{3/2}}/3\,,
\end{align}
and the corresponding horizon radii can be found to be
\begin{align}
    r_{E,N}&=r_U\sqrt{1\mp\sqrt{1-\rho^2}}\, .
\end{align}
Finally, we can use the relations between the roots of $f(r)$
 \begin{align}
& r_{--} + r_{-} + r_+ + r_{++}=0\,, \label{rel1}\\
 &r_{--}^2+ r_{-}^2 + r_+^2 + r_{++}^2=2l^2\,,\label{rel2}\\
 &r_{--} r_- + r_{--}r_+ + r_{--} r_{++} + r_{-}r_+ + r_{-}r_{++} + r_{+} r_{++} = - l^2\,, \label{rel3}
\end{align} 
  to find the following expressions for the other two  roots
\begin{align}
    r_-&=-\frac{1}{2}\left(r_{+}+r_{++}-\sqrt{2(2l^2-(r_+^2+r_{++}^2))-(r_++r_{++})^2}\right)\\
    r_{--}&=-(r_-+r_++r_{++})\,.
\end{align}
Note that \eqref{rel1} and \eqref{rel2} together yield a relation between the three positive roots and the de Sitter radius
\begin{equation}
    r_{-}^2+ r_+^2+r_{++}^2 + r_- r_+ + r_- r_{++}+ r_+ r_{++}=l^2\,.
\end{equation}

\subsection*{Horizon temperatures}
The radius $r_{\mathcal{O}}$ of the stationary free falling observer in $D=4$ is given by
\begin{equation}
    r_{\mathcal{O}}=a\left(1+\sqrt{\frac{m}{m_U}\frac{r_U^3}{a^3}-1}\,\,\right)\quad\text{where}\quad a\equiv r_U\sqrt{\rho\cos{\left[\frac{1}{3}\arccos{\left(\frac{m^2}{m_U^2}\rho^{-3}\right)}\right]}}\, ,
\end{equation}
and the black hole and cosmological horizon temperatures take the form 
\begin{equation}
     T_{+,++}=\frac{r_{++}-r_+}{4\pi l^2\sqrt{f(r_{\mathcal{O}})}}\left[2+\left(1+\frac{r_{+,++}}{r_{++,+}}\right)^2-\left(\frac{l}{r_{++,+}}\right)^2\right]\,.
\end{equation}
 For the  Nariai solution, where $r_+=r_{++}=r_{\mathcal{O}}=r_N$, these temperatures are the same and equal to
\begin{equation}
\label{t2eqi}
    T_N=\frac{\sqrt[4]{1-\rho^2}}{2\pi\,r_N}=\frac{\sqrt{3}}{\pi\,l\sqrt{2+2/\sqrt{1-\rho^2}}}\, .
\end{equation}
The other equilibrium solution for which $T_+=T_{++}$, but $r_+\neq r_{++}$, is the lukewarm solution,  which satisfies the relations $r_{L,+} +r_{L,++}=l$ and $m_{L}=2q_L$ in four dimensions. The horizon radii and temperature of the lukewarm solution  as a function of the charge are
\begin{equation}
    r_{{L,+,++}}=\frac{l}{2}\left(1\mp\sqrt{1-q/q_{LN}}\right),\qquad T_{L}=\frac{\sqrt{1-q/q_{LN}}}{2\pi\,l\sqrt{1-q^2/r_{\mathcal{O}}^2-3r_{\mathcal{O}}^2/l^2}}
    \, .
\end{equation}
Finally, for the lukewarm-Nariai solution we find 
\begin{equation}
  r_{LN} = \frac{l}{2}, \qquad m_{LN} =\frac{l}{2} , \qquad  q_{LN}= \frac{l}{4}, \qquad T_{LN}=\frac{1}{\sqrt{2}\pi \, l}\,.
\end{equation}


\section{Arbitrary dimensions}
In this appendix we collect some  cumbersome relations for the mass, charge and horizon radii of RNdS, which are valid in general dimensions. First, in arbitrary dimensions the relations \eqref{rel1}-\eqref{rel3} between the roots of the blackening factor $f(r)$ can be generalized to
 \begin{align}
\label{radrelation}
    \sum_{i=1}^{2(D-2)}r_i^n=\left\{
    \begin{array}{ll}
        0\quad\,\,\,\,\,\text{for $0<n\le D-2$\, odd}\\
        2l^n\,\,\,\,\,\,\text{for $0<n\le D-2$\, even\, }
    \end{array}\right. \qquad \text{and} \qquad
    \sum_{\substack{i,j=1 \\ i\neq j}}^{2(D-2)}r_i\,r_j=-l^2\,.
\end{align}
Next, by solving $f(r_+)= f(r_{++})=0$,   the mass and charge parameters can be expressed in terms of the outer horizon radius $r_+$, the cosmological horizon radius $r_{++}$ and de Sitter radius $l$  
\begin{equation}
\label{mq}
   m=\frac{r_+^{2(D-3)}(r_{+}^2-l^2)-r_{++}^{2(D-3)}(r_{++}^2-l^2)}{(r_{++}^{D-3}-r_+^{D-3})l^2}
    \,,\quad q^2=\frac{r_{++}^{D-3}(r_{++}^2-l^2)-r_+^{D-3}(r_{+}^2-l^2)}{(r_{++}^{3-D}-r_+^{3-D})l^2}\, .
\end{equation}
Further, for the lukewarm solution, solving $T_+=T_{++}$ (with the Bousso-Hawking normalization) yields the following expression for (the square of) the de Sitter radius  
\begin{align}
\label{lukerel}
   l^2=&\frac{1}{(D-3)\left(r_{L,++}^{D-2}-r_{L,+}^{D-2}\right)\left(r_{L,++}^{D-3}-r_{L,+}^{D-3}\right)^2}\Bigg [ (D-3)\left(r_{L,++}^{3(D-2)}-r_{L,+}^{3(D-2)}\right)\\
&-(D-2)\left(r_{L,++}^{D-4}-r_{L,+}^{D-4}\right)(r_{L,+}r_{L,++})^{D-1}-(D-1)\left(r_{L,++}^{D-2}-r_{L,+}^{D-2}\right) (r_{L,+}r_{L,++})^{D-2} \Bigg]\,. \nonumber
\end{align}
  By  plugging this into  equation \eqref{mq} one can obtain an equation for the ratio of the mass $m_L$ and charge $q_L$ in the lukewarm case    
\begin{align}\label{eq:mqel}
   \left(\frac{m_L}{q_L}\right)^2&=\frac{4(r_{L,+}+r_{L,++})}{l^2(D-3)\left(r_{L,++}^{D-2}-r_{L,+}^{D-2}\right)\left(r_{L,++}^{D-3}-r_{L,+}^{D-3}\right)^2}\,\,\, \times \\
  &\times\frac{\left[(D-3)\left(r_{L,++}^{2(D-2)}-r_{L,+}^{2(D-2)}\right)-(D-2)r_{L,+}r_{L,++}\left(r_{L,++}^{2(D-3)}-r_{L,+}^{2(D-3)}\right)\right]^2}{(D-3)\left(r_{L,++}^{D-1}-r_{L,+}^{D-1}\right)-(D-1)r_{L,+}r_{L,++}\left(r_{L,++}^{D-3}-r_{L,+}^{D-3}\right)} \nonumber \,.
\end{align}
In four dimensions this relation remarkably simplifies to $m_L = 2 q_L.$  The maximal values of the parameters $m_L$ and $q_L$ are set by the Nariai black hole. Solving $ m_L=m_N(\equiv m_{LN})$ and  $  q_L=q_N(\equiv q_{LN})$ yields in general dimensions
\begin{equation}
    \left(\frac{m_{LN}}{m_U}\right)^2=3^{D-3}(2D-5)^2\frac{(D-2)^{D-3}}{(3D-8)^{D-1}}\,,
    \qquad  \left(\frac{q_{LN}}{q_U}\right)^2=3^{D-3}\left(\frac{D-2}{3D-8}\right)^{D-2}
    \,,
\end{equation}
 such that    the horizon radius of the lukewarm-Nariai black hole is given by 
\begin{equation}
     r_{LN}^2=\frac{3(D-3)^2}{(D-1)(3D-8)}\,l^2\,.
\end{equation}
 \begin{figure}[t]
\centering
\begin{subfigure}{.5\textwidth}
  \centering
  \includegraphics[width=6.25cm]{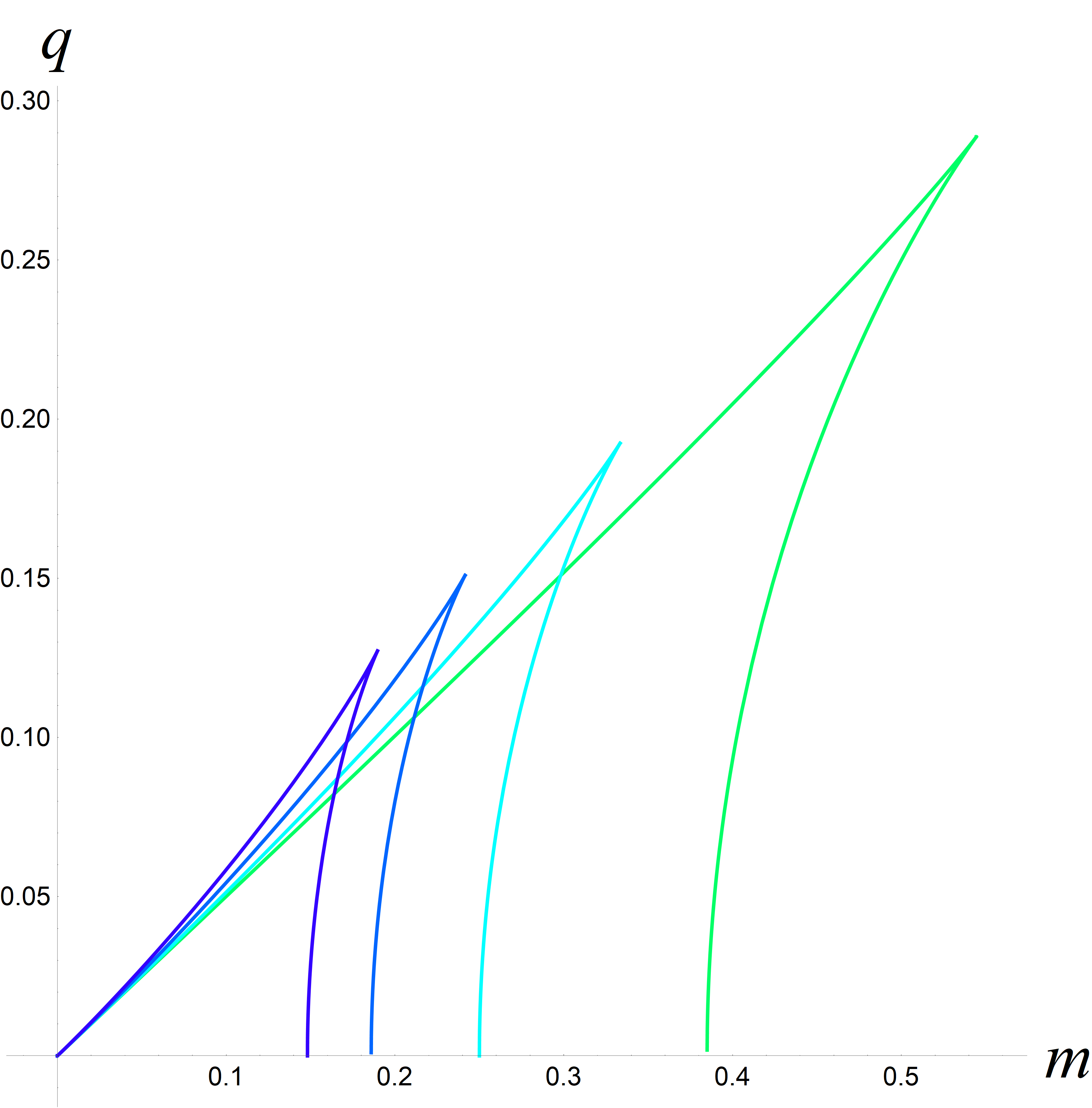}
  \caption*{a) Shark fins}
  \label{fig:c0}
\end{subfigure}%
\begin{subfigure}{.5\textwidth}
  \centering
  \includegraphics[width=6.75cm]{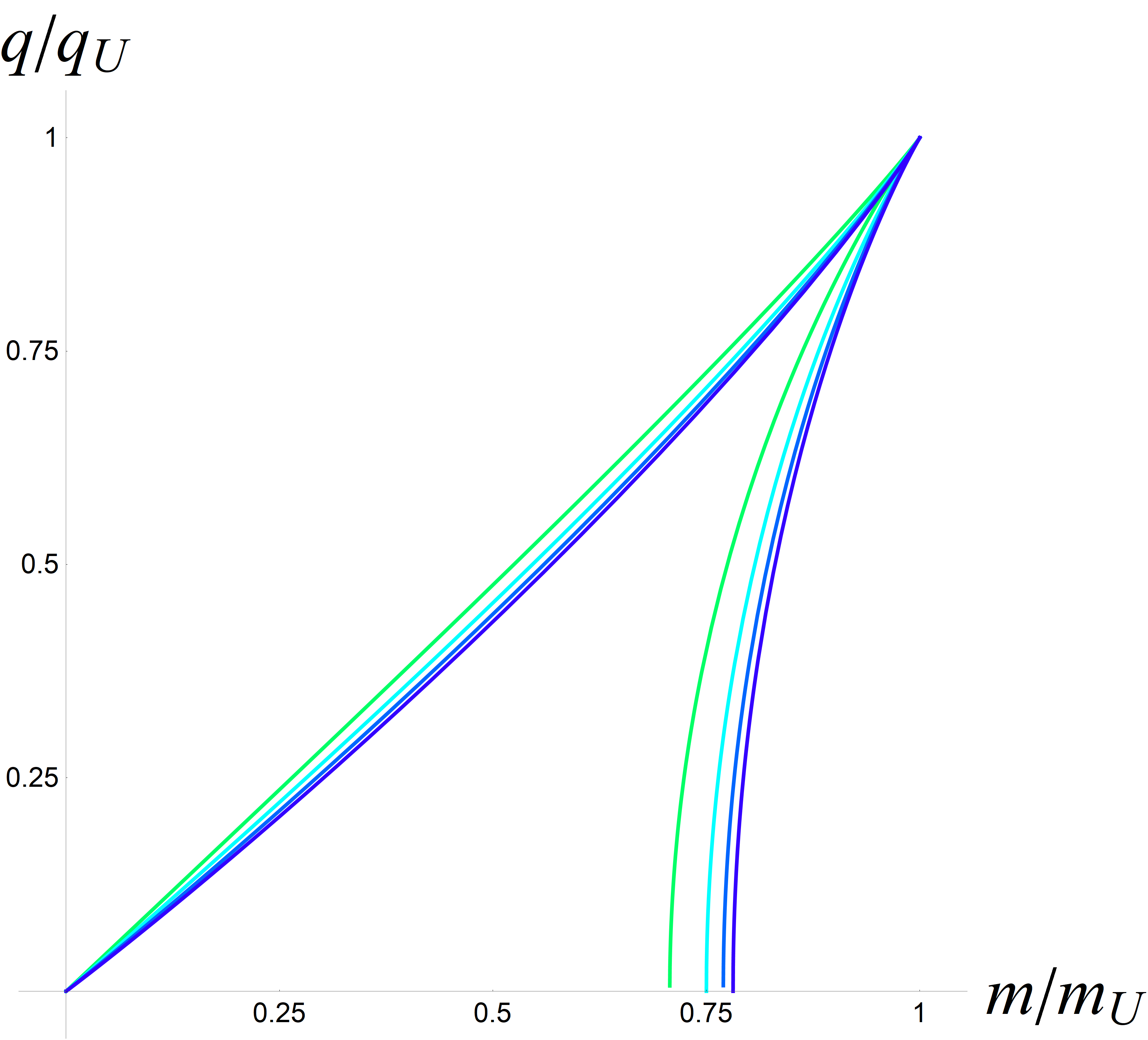}
  \caption*{b) Normalized shark fins}
  \label{fig:c736}
\end{subfigure}
\caption{Shark fin phase diagrams for $D=4$ (green), $D=5$ (turquoise), $D=6$ (blue) and $D=7$ (purple) obtained with \eqref{eq:mqbounds}. In the limit $D\rightarrow\infty$, the shark fins in the left panel get smaller and smaller, whereas the ones in the right panel converge to the values in  \eqref{eq:Narilim} and~\eqref{eq:Extlim}.} 
\label{fig:sharkcomp}
\end{figure}Finally, defining $\nu_N,\nu_E \in [0,1]$ via $r_N\equiv \nu_N\, r_U + (1-\nu_N)\, r_N^{0}$ and $r_E\equiv \nu_E^{1/D}\,r_U$, where $r_N^{0}= \sqrt{\frac{D-3}{D-1}} l$ stands for the neutral Nariai radius and the  radii  $r_{E,N}$ are related to $m_{E,N}$ and $q_{E,N}$  by \eqref{eq:mqbounds}, allows to obtain the limits of the ratios
 \begin{align}
     \frac{m_N}{m_U}&\xrightarrow{D\rightarrow\infty} \frac{1}{2} (1+\nu_N)\,e^{\frac{1}{2}(1-\nu_N)}\,,\qquad \frac{q_N}{q_U}\xrightarrow{D\rightarrow\infty} \sqrt{\nu_N}\,e^{\frac{1}{2}(1-\nu_N)} \label{eq:Narilim}\\
     \frac{m_E}{m_U}&\xrightarrow{D\rightarrow\infty} \nu_E (1-\ln\nu_E)\,,\qquad \qquad \,\,\,\frac{q_E}{q_U}\xrightarrow{D\rightarrow\infty} \nu_E\sqrt{1-2\ln\nu_E}\label{eq:Extlim}\,.
 \end{align}
These values set the boundaries of the normalized shark fin phase diagram in Figure \ref{fig:sharkcomp} in the $D\rightarrow\infty$ limit.

\bibliographystyle{JHEP}
\bibliography{refs}

\end{document}